\documentclass[structabstract]{aa}

\usepackage{graphicx}
\usepackage{subfigure}
\usepackage{amsmath}
\usepackage{amssymb}

\usepackage{natbib}
\bibpunct{(}{)}{;}{a}{}{,}
\newcommand{\vect}[1]{{\mathbf{#1}}}
\newcommand{\dd}{\mathrm{d}}

\begin{document}

\title{Numerical and semi-analytic core mass distributions in supersonic isothermal turbulence}
\author{Wolfram Schmidt\inst{1,2}
\and Sebastian A. W. Kern\inst{2,3}
\and Christoph Federrath\inst{3,4}
\and Ralf S. Klessen\inst{4}
}

\authorrunning{W. Schmidt et al.}
\titlerunning{Core mass distributions in supersonic isothermal turbulence}

\institute{
	Institut f\"ur Astrophysik, Universit\"at G\"ottingen, Friedrich-Hund-Platz 1,
	D-37077 G\"ottingen, Germany\\
              	\email{schmidt@astro.physik.uni-goettingen.de}
               \and
	Lehrstuhl f\"ur Astronomie, Institut f\"ur Theoretische Physik und Astrophysik, Universit\"at W\"urzburg, Am Hubland, D--97074 W\"urzburg, Germany
	\and
	Max-Planck-Institut f\"ur Astronomie, K\"onigstuhl 17, D--69117 Heidelberg, Germany\\
                \email{kern@mpia-hd.mpg.de}
	\and
	Zentrum f\"ur Astronomie, Institut f\"ur Theoretische Astrophysik, Universit\"at Heidelberg, Albert-Ueberle-Str. 2, D--69120 Heidelberg, Germany\\
            	\email{chfeder@ita.uni-heidelberg.de}
}

\date{DRAFT: \today}

\abstract
{Supersonic turbulence in the interstellar medium plays an important role in the formation of stars. The origin of this observed turbulence and its impact on the stellar initial mass function (IMF) still remains an open question.}
{We investigate the influence of the turbulence forcing on the mass distributions of gravitationally unstable cores in simulations of isothermal supersonic turbulence.
}
{Data from two sets of non-selfgravitating hydrodynamic FLASH3 simulations with external stochastic forcing are analysed, each with static grid resolutions of $256^3$, $512^3$ and $1024^3$ grid points. The first set applies solenoidal (divergence-free) forcing, while the second set uses purely compressive (curl-free) forcing to excite turbulent motions. From the resulting density field, we compute the mass distribution of gravitationally unstable cores by means of a clump-finding algorithm. Using the time-averaged probability density functions of the mass density, semi-analytic mass distributions are calculated from analytical theories. We apply stability criteria that are based on the Bonnor-Ebert mass resulting from the thermal pressure and from the sum of thermal and turbulent pressure. 
}
{Although there are uncertainties in the application of the clump-finding algorithm, we find systematic differences in the mass distributions obtained from solenoidal and compressive forcing. Compressive forcing produces a shallower slope in the high-mass power-law regime compared to solenoidal forcing. The mass distributions also depend on the Jeans length resulting from the choice of the mass in the computational box, which is freely scalable for non-selfgravitating isothermal turbulence. If the Jeans length corresponding to the density peaks is less than the grid cell size, the distributions obtained by clump-finding show a strong resolution dependence. Provided that all cores are numerically resolved and most cores are small compared to the length scale of the forcing, the normalised core mass distributions are close to the semi-analytic models.}
{The driving mechanism of turbulence has a potential impact on the shape of the core mass function. Especially for the high-mass tails, the Hennebelle-Chabrier theory implies that the additional support due to turbulent pressure is important.}


\maketitle

\section{Introduction}\label{sc:introduction}

The observed supersonic turbulence in the interstellar medium (ISM) plays an important role in the process of star formation \citep[e.g., ][]{MacLowKlessen2004,ElmegreenScalo2004,ScaloElmegreen2004, BallKless07,McKeeOstriker2007}. The origin of this turbulence and the characteristics of different turbulence driving mechanisms are still a matter of debate. Due to the ability to counterbalance gravity globally and to provoke gravitational collapse locally turbulence is expected to have a strong impact on the shape of the stellar Initial Mass Function (IMF). The high-mass end ($m\gtrsim 0.5\, \mathrm{M_{\sun}}$) is typically observed to exhibit a power law of the form $\mathcal{N}(m)\mathrm{d}m \propto m^{-\alpha}\mathrm{d}m$ \citep[e.g.,][]{Salpeter1955, Kroupa2001, Chabrier2003, Lada2003}. $\mathcal{N}(m)$ is the number of stars in the linear mass interval $\mathrm{d}m$ with \citet{Salpeter1955} power-law index $\alpha=2.35$. Observations of dense cores embedded in star forming regions of turbulent giant molecular clouds show a similar but slightly shallower power-law distribution of mass with an exponent $\alpha$ in the range of $1.5-2.5$ \citep{ElmegreenFalagrone1996,Motte1998}. The origin of the IMF and the Salpeter power law for the high-mass tail are still under debate but there is a possible connection between the core mass function (CMF) and the IMF \citetext{\citealp{Motte1998, TestiSargent1998, WilliamsBlitz2000, JohnstoneEtAl2000, JohnstoneEtAl2001, Johnstone2006, Nutter2007}; \citealp*{AlvesLombardi2007}; see however, the critical discussion by \citealp*{ClarkKlessen2007}}. \citet{PadoanNordlund2002} proposed a theoretical explanation for the CMF/IMF based on scaling properties of turbulence and the Jeans criterion for gravitational instability. Combining the formalism of
\citet{PressSchech74} for cosmological structure formation with the notion of a turbulent Jeans mass, a new analytic theory of the IMF was formulated by \citet{HennebelleChabrier2008}. 

Numerical simulations of turbulent molecular clouds reported in the literature \citep[for instance,][]{KritsukEtAl2007,HenBan08,BanVaz08,SchmidtEtAl2009,FederDuv09} inject the turbulent energy 
via different methods, but apart from an early study by \citet{Kless01}, there has been no systematic analysis of how different driving schemes affect the CMF. Possible physical mechanisms for maintaining
ISM turbulence range from stellar feedback (supernovae, outflows, ionizing radiation) to large-scale dynamical instabilities in the Galaxy \citep{ElmegreenScalo2004,MacLowKlessen2004,NormanFerrara1996},
In this work we investigate the influence of compressibility of the turbulence forcing on the mass spectra of gravitationally unstable cores. 

We apply a clump-finding algorithm to the density fields from recent simulations of supersonic isothermal gas, using the FLASH3 code to solve the equations of hydrodynamics on static grids with resolutions of $256^3$, $512^3$ and $1024^3$ grid points and periodic boundary conditions \citep{Federrath2008,FederDuv09,Federrath2009}. The turbulent energy is continuously injected by a specific force $\vect{f}$ on length scales corresponding to wavenumbers $1<k<3$, where $k$ is normalised by $2\pi/X$ for a box of size $X$. Each component of this force is modelled by an Ornstein-Uhlenbeck process \citep{EswaranPope1988,SchmidtEtAl2006,SchmidtEtAl2009,FederDuv09}. We define the integral scale $L$ by the mean wavelength of the forcing, i.~e., $L=X/2$. The relative importance of solenoidal and compressive modes of the stochastic forcing is set by the parameter $\zeta \in [0,1]$. The simulations represent two extreme cases: Purely solenoidal forcing ($\nabla\cdot\vect{f}=0$, $\zeta=1$) on the one hand and purely compressive forcing ($\nabla\times\vect{f}=0$, $\zeta=0$) on the other driven to a RMS Mach number of ${\fam=2 M}_{\mathrm{rms}}\approx 5.5$. Turbulence becomes statistically stationary for $t \geq 2\ \mathcal{T}$, where $\mathcal{T}$ is the dynamical timescale. In order to take the intermittent nature of the density field into account we analyse each of the $81$ density snapshots available in the time interval $2\,\mathcal{T} \leq t \leq 10\,\mathcal{T}$ to compute time-averaged core statistics. For a detailed description of the simulation setup and numerical techniques see \citet[][]{FederDuv09}. Recent results of isothermal turbulence simulations indicate a strong dependence of the density field statistics on the compressibility of the stochastic forcing \citep{Federrath2008, SchmidtEtAl2009,FederDuv09,Federrath2009}. Due to this effect it is a reasonable assumption that the resulting core mass distributions should show different properties for different turbulence driving mechanisms. Furthermore, the datasets enable us to study in which way the effective resolution of the simulation affects the resulting mass distributions. 

This article is structured as follows. In the next section, we will review the analytical theories for the
CMF by \citet{PadoanNordlund2002} and \citet{HennebelleChabrier2008}. For the sake of clarity, we express the main results of these theories in a consistent notation. In Sect.~\ref{sc:scaling}, we discuss the parameters determining the global mass scale, which plays a key role in computations of the core mass functions. The clump-finding algorithm is briefly described in Sect.~\ref{sc:statistics}, and the resulting mass distributions are presented. Particular emphasis is put on the question of numerical convergence. The distributions obtained by clump-finding are compared to semi-analytic computations of the CMF in Sect.~\ref{sc:comparison}, and the results are summarized and discussed in 
Section~\ref{sc:conclusion}.

\section{Analytical theories for the core mass function}

In this work, we follow \citet{HennebelleChabrier2008} and write $\mathcal{N}(M)=\dd N/\dd M$, where $N(M)$ is the cumulative number of cores with masses below $M$. The function $\mathcal{N}(M)$ is also called the mass spectrum. Because of the wide range of different masses, it is useful to define the CMF by the number of cores per logarithmic mass interval. To that end, we define
two dimensionless mass variables:
\begin{itemize}
\item $m:=M/M_{\sun}$ is the mass relative to the solar mass. 
\item $\tilde{M}:=M/M_{\mathrm{J}}^{0}$ is the mass relative to the thermal Jeans mass
at the mean density $\bar{\rho}$.
\end{itemize}
While $m$ is useful to discuss the CMF in an observational context, $\tilde{M}$ is the most
natural choice for theoretical considerations. 
The linear mass distribution is related to the distributions with respect to $\log m$ or $\log\tilde{M}$ as follows:\footnote{
In this article, $\log$ denotes the natural logarithm.}
\begin{equation}
	\label{eq:number_dens}
	\mathcal{N}(M) =
	(M_{\sun}m)^{-1}\frac{\dd N}{\dd\log m} =
	(M_{\mathrm{J}}^{0}\tilde{M})^{-1}\frac{\dd N}{\dd\log\tilde{M}}.
\end{equation}
Integration of over $M$ yields the total number of cores, 
\begin{equation}
\label{eq:number_tot}
\begin{split}
	N_{\mathrm{tot}} =& \int_{0}^{\infty}\mathcal{N}(M)\,\dd M = \\ 
 		& \int_{-\infty}^{\infty}\frac{\dd N}{\dd\log m}\,\dd\,\log m
		  = \int_{-\infty}^{\infty}\frac{\dd N}{\dd\log\tilde{M}}\,\dd\,\log\tilde{M}.
\end{split}
\end{equation}

\subsection{Padoan-Nordlund theory}

\label{sc:pn_theory}

\citet{PadoanNordlund2002} based their analytical approach on the following assumptions:
The core size is comparable to the thickness of isothermal postshock gas, the turbulent velocity fluctuations follow a power law, and the minimal mass that is unstable against gravitational collapse is given by the thermal Jeans mass. They showed that the number of cores per mass interval is given by an expression of the form
\begin{equation}
	\label{eq:PN_mass_spect_org}
	\mathcal{N}(M) \propto 
	M^{-(x+1)}\int_{0}^{M}\mathrm{pdf}_{\mathrm{J}}(M')\,\dd M',
\end{equation}
where the parameter $x$ depends on the scaling properties of turbulence. Integration of the probability density function $\mathrm{pdf}_{\mathrm{J}}(M')$ of the Jeans mass for a given distribution of the mass density yields the fraction of cores with masses $M'\le M$ that are unstable according to the Jeans criterion, $M'\propto\rho^{-1/2}$. For cores of sufficiently high mass, this integral asymptotically approaches a constant. For this reason, the high-mass tail of $\mathcal{N}(M) $ obeys a power law with exponent $-(x+1)$.  

Equation~(\ref{eq:PN_mass_spect_org}) can be written in terms of the probability density function
$\mathrm{pdf}(\delta)$ of the logarithmic density fluctuation $\delta=\log(\rho/\bar{\rho})$.
Since the Jeans stability criterion implies $\log\tilde{M}=-\frac{1}{2}\delta$,
we have $\mathrm{pdf}_{\mathrm{J}}(M')=-2\mathrm{pdf}(\delta)/M'$, and it follows that
\begin{equation}
\label{eq:PN_mass_spect}
	\frac{\dd N}{\dd\log\tilde{M}} \propto 
	\tilde{M}^{-x}\int_{-2\log\tilde{M}}^{\infty}\mathrm{pdf}(\delta)\,\dd\delta.
\end{equation}
Note that the upper bound on the core mass corresponds to a lower bound on the density
fluctuation. The distribution~(\ref{eq:PN_mass_spect}) is invariant under a change of the global scales. Obviously, this cannot be the physical CMF if $M_{\mathrm{J}}^{0}\gtrsim M_{\mathrm{tot}}$, where $M_{\mathrm{tot}}:=(2L)^{3}\bar{\rho}$ is the total mass in the computational domain (also see the discussion of the mass scale in Sect.~\ref{sc:scaling}). For this reason, the Padoan-Nordlund theory has to be considered as asymptotic description for $M_{\mathrm{J}}^{0}\ll M_{\mathrm{tot}}$. The analytic calculation of the mass spectrum is straightforward if the probability density function of the mass-density is lognormal, i.~e.,
\begin{equation}
	\mathrm{pdf}(\delta)=
	\frac{1}{\sqrt{2\pi\sigma^{2}}}\exp\left[-\frac{(\delta-\bar{\delta})^{2}}{2\sigma^{2} }\right].
\end{equation}
where $\bar{\delta}=-\sigma_{0}^{2}/2$.

Originally, the power-law exponent $x$ was obtained from the jump conditions for shocks in magnetohydrodynamic turbulence, and it was shown that the resulting slope of the high-mass tail is close to the Salpeter slope $x=\alpha-1=1.35$ \cite[see][]{PadoanNordlund2002}. \citet{Padoan2007b}, hereafter referred to as PN07, determined $x$ also for isothermal hydrodynamic (HD) turbulence: 
\begin{equation}\label{eq:xslope}
x_{\mathrm{HD}}=\frac{3}{5-2\beta},
\end{equation}
where $\beta$ is the slope of the turbulence energy spectrum. Using data from a driven isothermal HD turbulence simulation, PN07 calculated numerical mass distributions by means of clump-finding for purely thermal support. The spectral index $\beta=1.9$ \citep{KritsukEtAl2007} implies $x_{\mathrm{HD}}=2.5$ for the slope of the high-mass tail. This value recovers well the general trend of their numerical mass  distributions (see Fig. 2 and Fig. 3 in PN07). However, they did not perform time-averaging for their mass distributions, hence no estimate of the statistical significance of their results was provided. It was concluded that the high-mass tail is generally too stiff for HD turbulence to
be consistent with the observed CMF.

\subsection{Hennebelle-Chabrier theory}

\label{sc:hc_theory}

The theory of \citet{HennebelleChabrier2008} provides a general framework for mass distributions
of cores on the basis of the Press-Schechter statistical formalism. Other than in the
theory by  \citet{PadoanNordlund2002}, there is an inherent notion of a core scale $R$ that is connected to its mass. Assuming a log-normal distribution of the gas density, analytic formulae for the mass distribution $\mathcal{N}(M)$ can be derived. In the simplest case, $M$ is given by the thermal Jeans mass for a given density. While the Padoan-Nordlund theory only uses thermal support, the Hennebelle-Chabrier theory considers a combination of thermal and turbulent support. On the one hand, turbulence is promoting star formation through the broadening of the density PDF, while, on other hand, it is also quenching star formation through the scale-dependent support it exerts on cores.

Applying an approximation that excludes the largest cores, it is possible to generalise their approach to arbitrary PDFs of the mass-density. For cores on a length scale $R$ comparable to the integral scale $L$, the mass distribution depends on the derivative of $\mathrm{pdf}(\delta)$ with respect to $R$, which, in turn, depends on the slope of the
spectrum of the density fluctuations. It would be worthwhile to investigate the influence of
the spectral properties of the density field on the CMF. Because of the difficulties in the numerical determination of the derivative of the $\delta$-PDF, however, we restrict our investigation to
length scales $R\ll L$, for which the general form of the mass spectrum \citep[equation (33) in][]{HennebelleChabrier2008} simplifies to \
\begin{equation}
\label{eq:HC_mass_spect}
	\mathcal{N}(M) \simeq
	-\bar{\rho}\left(M\frac{\dd M}{\dd R}\right)^{-1}
	\frac{\dd\delta}{\dd R}\exp(\delta)\mathrm{pdf}(\delta).
\end{equation}

For a comparison with the Padoan-Nordlund theory, the case of purely thermal support is treated as follows. The core mass $M=M_{\mathrm{J}}^{0}\tilde{M}$ is related to the logarithmic density fluctuation $\delta$ by the Jeans criterion for stability against gravitational collapse, and $R$ is given by the thermal Jeans length. Hence,
\begin{equation}
\label{eq:mass_dens_thermal}
\tilde{M}=\tilde{R}=\exp(-\delta/2).
\end{equation}
The dimensionless scale parameter is defined by $\tilde{R}:=R/\lambda_{\mathrm{J}}^{0}$, where $\lambda_{\mathrm{J}}^{0}:=a_{\mathrm{J}}^{1/3}c_{\mathrm{s}}/(G\bar{\rho})^{1/2}$ is the thermal Jeans length for the mean density ($c_{s}$ is the isothermal speed of sound, $G$ the gravitational
constant, and $a_{\mathrm{J}}$ a geometry factor). 
Substituting the inverse mass-density relation~(\ref{eq:mass_dens_thermal}), 
it follows from equation~(\ref{eq:HC_mass_spect}) that
\begin{equation}
\label{eq:HC_mass_spect_log_therm}
	\frac{\dd N}{\dd\log\tilde{M}}=
	-\frac{\bar{\rho}}{M_{\mathrm{J}}^{0}}\tilde{M}^{-3}\mathrm{pdf}(-2\log\tilde{M}).
\end{equation}
We note that this expression for the CMF is fully determined by the lower bound of the integral in equation~(\ref{eq:PN_mass_spect}), and, generally, the high-mass tail does not obey a power law.

If turbulence pressure contributes to the support against gravitational collapse, an implicit relation between the gravitationally unstable mass and the density fluctuation results:
\begin{align}
\label{eq:mass_dens}
\tilde{M} = \tilde{R}\left(1+\mathcal{M}_{\ast}^{2}\tilde{R}^{2\eta}\right),
& \quad \delta = \log\left(\frac{1+\mathcal{M}_{\ast}^{2}\tilde{R}^{2\eta}}{\tilde{R}^{2}}\right).
\end{align}
The intensity of turbulence is specified by the Mach number $\mathcal{M}_{\ast}$ of turbulent velocity fluctuations on the length scale $\lambda_{\mathrm{J}}^{0}$,
\begin{equation}
\label{eq:mach_star}
	\mathcal{M}_{\ast} := \frac{\mathcal{M}_{\mathrm{rms}}}{\sqrt{3}}
	\left(\frac{\lambda_{\mathrm{J}}^{0}}{L}\right)^{\eta},
\end{equation}
where $\eta=(\beta-1)/2$, and $\beta$ is the slope of the turbulence energy spectrum function in the inertial subrange. Since the turbulent pressure on the length scale $\lambda_{\mathrm{J}}^{0}$ equals $\rho(\mathcal{M}_{\mathrm{\ast}}c_{\mathrm{s}})^{2}$, the parameter $\mathcal{M}_{\ast}$ measures the relative significance of turbulent vs.\ thermal support for cores of size $\sim \lambda_{\mathrm{J}}^{0}$. The turbulent Mach number on the length scale $R$ is given
by $\mathcal{M}_{\ast}\tilde{R}^{\eta}$. Note that $\mathcal{M}_{\ast}\tilde{R}^{\eta}$ is about unity if $R$ is close to the sonic length scale \citep{SchmidtEtAl2009,FederDuv09}.
The CMF including turbulent support is given by
\begin{equation}
\label{eq:HC_mass_spect_log}
\begin{split}
	\frac{\dd N}{\dd\log\tilde{M}}=&
	\frac{2\bar{\rho}}{M_{\mathrm{J}}^{0}}
	\frac{1+(1-\eta)\mathcal{M}_{\ast}^{2}\tilde{R}^{2\eta}}{\tilde{R}^{3}[1+(2\eta+1)\mathcal{M}_{\ast}^{2}\tilde{R}^{2\eta}]}\\
	&\times\mathrm{pdf}\left[\log(\tilde{M}/\tilde{R}^{3})\right],
\end{split}
\end{equation}
where $\tilde{R}$ is numerically determined by the inversion of equation~(\ref{eq:mass_dens}) for a given value of $\tilde{M}$. 

\section{Clump-finding and scaling}\label{sc:scaling}

For the calculation of the mass distributions we applied the implementation of the clump-finding algorithm by PN07. As a first step the algorithm divides the three-dimensional density field into $k$ discrete density levels $\rho_{i}$ based upon the settings of the input parameters $\rho_{\mathrm{min}}$ and $f=\rho_{i+1}/\rho_{i}$ with  $i=1\ldots k$. The former parameter defines the minimum density level in units of the average density $\bar{\rho}$ while the latter defines the spacing between two adjacent levels. In the next step each density level is scanned for regions of connected cells with density values higher than the current density level. A connected region
is counted as an object (a core) if its mass exceeds the local Bonnor-Ebert mass \citep{Bonnor1956, Ebert1955}. The ratio of the Bonnor-Ebert mass to the solar mass is given by
\begin{equation}\label{eq:bonnorebert}
\begin{split}
&m_{\mathrm{BE}}=\frac{1.18 c_{\mathrm{s}}^3}{M_{\sun}G^{3/2}\rho^{1/2}}=\\
&= 3.19\, \left(\frac{\mu}{2.5}\right)^{-2}\left(\frac{n}{1000\, \mathrm{cm^{-3}}}\right)^{-1/2}\left(\frac{T}{10\,\mathrm{K}}\right)^{3/2},
\end{split}
\end{equation}
where the gravitational constant $G=6.67\times 10^{-8}\, \mathrm{cm^3\, g^{-1}\, s^{-2}}$, $c_{\mathrm{s}}$ is the isothermal sound speed, $T$ is the temperature, and $\mu$ is the mean molecular weight of the gas. The number density $n=\rho/(\mu m_{\mathrm{H}})$, where $m_{\mathrm{H}}$ is the mass of the hydrogen atom, is averaged over the connected region. In the last step, each object which can be split into two or more objects is rejected. For a more detailed description of the clump-finding method, see PN07.

For the clump-finding analysis, it is crucial to consider the degrees of freedom in the scaling of 
isothermal, non-self-gravitating turbulence. For a given temperature $T$ and box size $L$,
the hydrodynamical scales of the system are fixed, and the forcing magnitude determines
the RMS Mach number $\mathcal{M}_{\mathrm{rms}}$ of turbulence in the statistically stationary state. If self-gravity is not included, the flow properties are invariant with respect to the choice
of the mean mass density $\bar{\rho}$. This is palpable if the compressible
Euler equations are written in terms of the logarithmic density $\delta=\ln(\rho/\bar{\rho})$ \citep[see, for instance,][]{FederDuv09}. Therefore, the mass scale is arbitrary. 

On the other hand, if gravitationally unstable cores are identified via postprocessing, then the value of $m_{\mathrm{BE}}$~(\ref{eq:bonnorebert}) resulting from the chosen values of $n$ and $T$ fixes the mass scale. In this paper, we characterize the mass scale by the total number $N_{\mathrm{BE}}$ of Bonnor-Ebert masses with respect to the mean density contained in the simulation box:
\begin{equation}\label{eq:nj}
\begin{split}
&N_{\mathrm{BE}} =\frac{m_{\mathrm{tot}}}{m_{\mathrm{BE}}(\bar{\rho})}= 10^2\ \mathrm{M_{\sun}} \left(\frac{\mu}{2.5}\right)^3\left(\frac{\bar{n}}{1000\  \mathrm{cm^{-3}}}\right)^{3/2}\times \\
&\times \left(\frac{T}{10\ \mathrm{K}}\right)^{-3/2}\left(\frac{2L}{1.782\ \mathrm{pc}}\right)^{3},
\end{split}
\end{equation}
where $m_{\mathrm{BE}}(\bar{\rho})$ is the Bonnor-Ebert mass for the mean density $\bar{\rho}$, and $\bar{n}=\bar{\rho}/(\mu m_{\mathrm{H}})$. The ratio of the thermal Jeans length with the geometry factor $a_{\mathrm{J}}=1.18$ corresponding to the definition of the Bonnor-Ebert mass to the integral length is given by
\begin{equation}
\frac{\lambda_{\mathrm{J}}^{0}}{L}=\frac{2}{N_{\mathrm{BE}}^{1/3}}.
\end{equation}

The Hennebelle-Chabrier theory accounts for the dependence of the core statistics
on the parameter $N_{\mathrm{BE}}$ via the factor $\bar{\rho}/M_{\mathrm{J}}^{0}\sim N_{\mathrm{BE}}^3/(2L)^3$ in the distributions (\ref{eq:HC_mass_spect_log_therm}) and~(\ref{eq:HC_mass_spect_log}). Apart from that, variations of the mass scale correspond to changes of the scale-dependent Mach number,
\begin{equation}
\label{eq:mach_star_be}
	\mathcal{M}_{\ast} = 
	\frac{\mathcal{M}_{\mathrm{rms}}}{\sqrt{3}}
	\left(\frac{N_{\mathrm{BE}}}{8}\right)^{-\eta/3},
\end{equation}
where $\mathcal{M}_{\ast}$ characterizes the ratio of turbulent pressure to thermal
pressure on the length scale $\lambda_{\mathrm{J}}^{0}$ (see eqn.~\ref{eq:mach_star}).
In order to include the influence of the turbulent pressure on the stability against gravitational collapse
in the clump-finding tool, we modified the stability criterion by replacing the thermal speed of sound in the definition of the Bonnor-Ebert mass with an effective speed of sound \cite[see, for instance,][]{BonaHey87}:
\begin{equation}\label{eq:cseff}
c_{\mathrm{eff}}^{2} = c_{\mathrm{s}}^{2} + \frac{1}{3}\sigma_{\mathrm{core}}^{2}(v).
\end{equation}
The turbulent velocity dispersion $\sigma_{\mathrm{core}}(v)$ is computed from the mass-weighted RMS velocity fluctuation with respect to the centre-of-mass velocity for a particular core. 

The physical scales that were chosen by PN07 are $n = 10^4\;\mathrm{cm^{-3}}$, $X = 2L = 6\;\mathrm{pc}$ and $T = 10\;\mathrm{K}$. In this case, $N_{\mathrm{BE}}\approx 1.2\times 10^{5}$ and $\lambda_{\mathrm{J}}^{0}/L\approx 0.04$. For the simulations with the highest
resolution, we have $\lambda_{\mathrm{J}}^{0}\approx 20.8\Delta$, where $\Delta$ is the
size of the grid cells. As we shall see in the next Section, this introduces a serious resolution issue, because many cores extend only over a few cells for this parameter set. A further difficulty is that the above parameters are atypical for observed molecular cloud properties.
According to the \citet{Larson85} relation, $\bar{n}\approx 3000\;\mathrm{cm^{-3}}(L/1\,\mathrm{pc})^{-1}$ \citep[see also][]{FalPug92,HeyKraw09}. This suggests that the assumed forcing scale $L=X/2=3\;\mathrm{pc}$
is too large for molecular clouds of mean number density $\bar{n}=10^4\;\mathrm{cm^{-3}}$.
In order to analyse the impact of the physical scales, we re-computed the core mass distributions for $L = 0.6\;\mathrm{pc}$, without changing the temperature and the mean number density. For these parameters, the number density is consistent with the Larson relation within a factor of two, which is reasonable given the scattering of observed molecular cloud properties \citep{FalPug92}. In this case, $N_{\mathrm{BE}}=10^{3}$, and $\lambda_{\mathrm{J}}^{0}/L\approx 0.2$. We also note, however, that the validity of Larson-type relations for the mass density was questioned by \citet{BallMacLow02}. 

\begin{figure*}[t]
\subfigure[solenoidal forcing]{\label{fg:solfdependence}\includegraphics[width=0.49\textwidth]{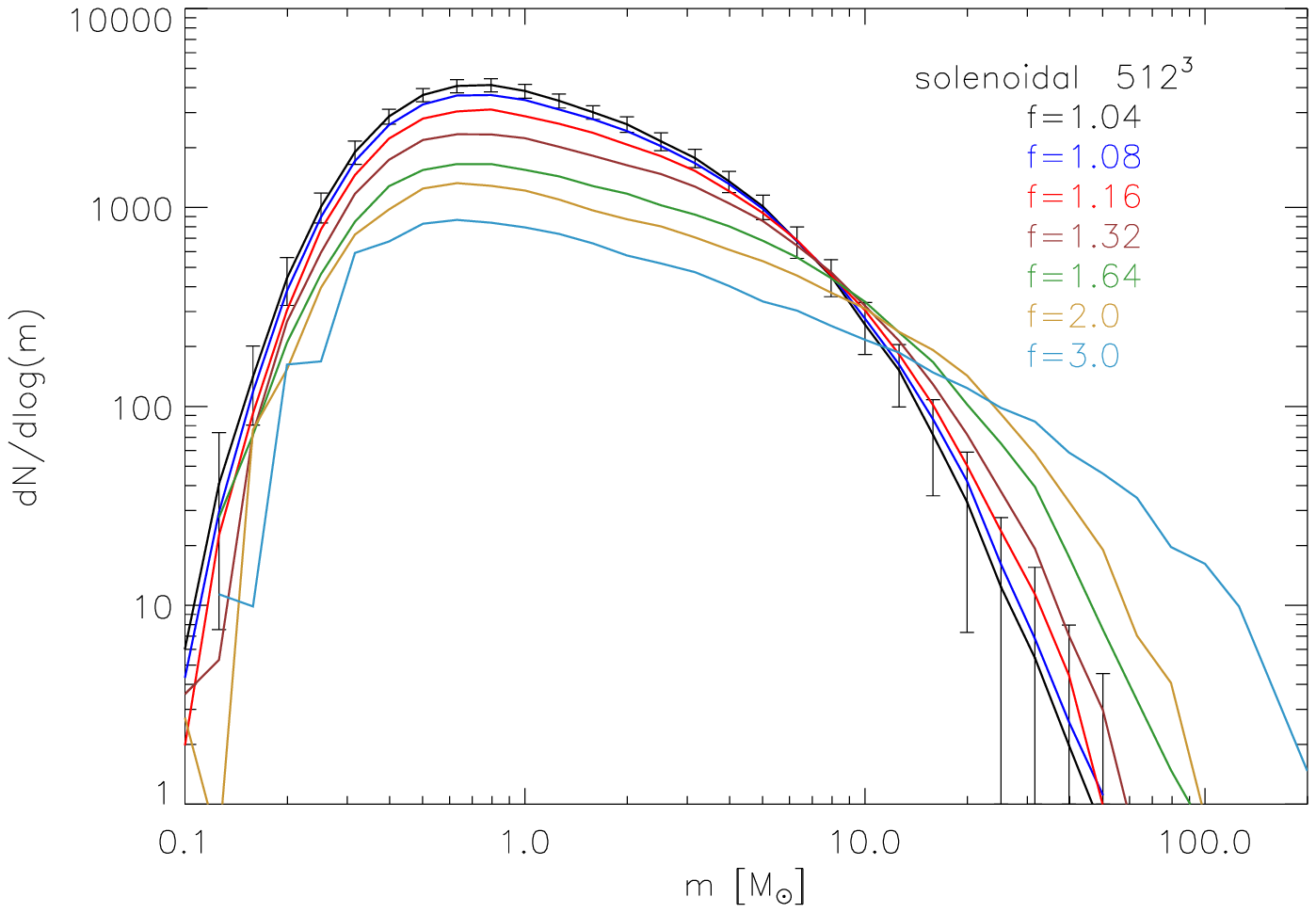}}\hfill
\subfigure[compressive forcing]{\label{fg:dilfdependence}\includegraphics[width=0.49\textwidth]{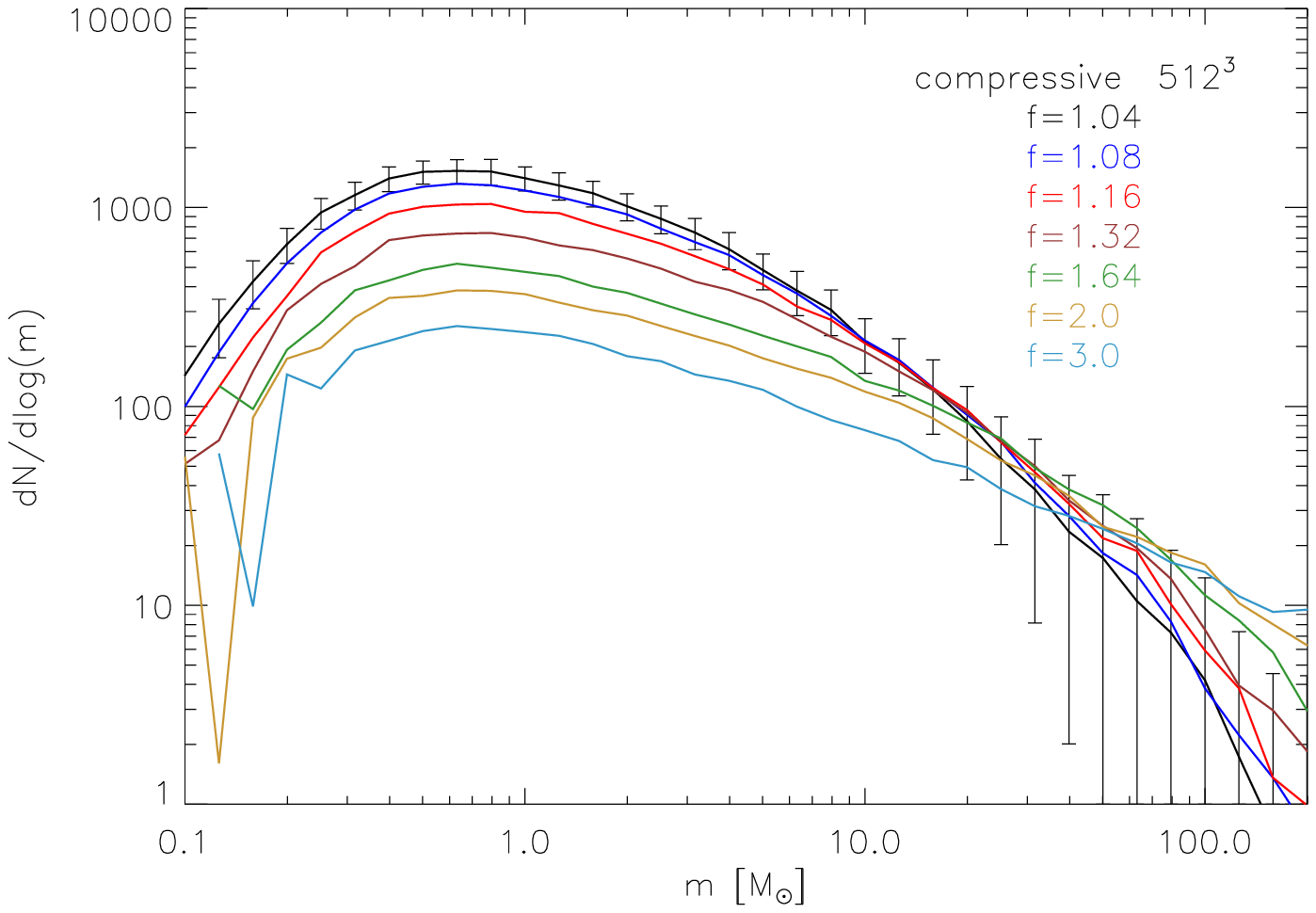}}
\caption{The core mass distributions for the $512^3$ solenoidal \subref{fg:solfdependence} and compressive \subref{fg:dilfdependence} simulations as a function of the clump-finding algorithm parameter $f$ (increasing from top curve to bottom curve) which sets the relative spacing between two adjacent density levels. Error bars contain the 1$\sigma$ temporal fluctuations and are only indicated for $f=1.04$ for the sake of clarity.}
  \label{fg:fdependence}
\end{figure*}

\section{Core statistics}\label{sc:statistics}

We begin with a detailed analysis of the mass distributions obtained for
purely thermal support and the PN07 mass scale corresponding to
$N_{\mathrm{BE}}\approx 1.2\times 10^{5}$. In particular, the tuning of
clump-finding algorithm and influence of numerical resolution are
considered in Section \ref{sc:clumpfind_tuning} and \ref{sc:num_res}. Finally, the mass distributions for a smaller number of
Bonner-Ebert masses in the box, $N_{\mathrm{BE}}\approx 1.0\times 10^{3}$, as
well as the influence of turbulent pressure are discussed in Section \ref{sc:turbpressure_massscale}.

\subsection{Tuning of the clump-finding algorithm}
\label{sc:clumpfind_tuning}

As described in section \ref{sc:scaling} the clump-finding routine needs two
technical input parameters. The relative spacing $f$ between two adjacent density levels and the minimum
density level $\rho_{\mathrm{min}}$ above which the datacube is divided into
discrete levels. The influence of these parameters was already described in
PN07. For the minimum density level we choose the mean density $\rho_{0}$ of
the simulation box since we observe no significant differences in the
resulting core mass distributions for values of $\rho_{\mathrm{min}} < \rho_{0}$ which
was also found in PN07. The influence of the level spacing $f$ on the
resulting mass distributions is shown in Fig. \ref{fg:fdependence}.
The slope of the high-mass range and the total number of cores is very
sensitive to the chosen spacing of the density levels. For larger $f$-values
the algorithm detects less but more massive cores. For decreasing spacing between density levels and therefore finer ``scanning'' of the datacube, the massive cores are split into a higher
number of lower mass cores which leads to a steepening of the high-mass tail
of the core mass distributions. For $f=1.04$, $1.08$ and $1.16$, this effect
lies within the temporal fluctuations of the mass distributions as indicated
by the error bars in Fig. \ref{fg:fdependence}. 
PN07 also concluded that the mass distributions are basically converged for $f \leq 1.16$. 
For the calculations in the present study, we chose a value of $f=1.04$. 
Nevertheless, comparing our numerical mass distributions to
the observed IMF/CMF should be done with caution, especially, in the high-mass range.
Note that the parameter dependency of core mass distributions is a common
problem for \textit{all} clump-finding methods in the
literature. \citet{SmithClark2008} applied different clump-finding methods to data
of SPH simulations of molecular clouds with the result that the definition of an individual core and its mass depends strongly on the method
and the parameter settings \citep[see also ][]{KlessenBurkert2000, Kless01}. In the observers community the clump-finding routines CLUMPFIND
\citep{WilliamsDeGeus1994} and GAUSSCLUMPS \citep{StutzkiGuesten1990} are commonly
used to decompose position-position-velocity information of molecular clouds
into individual cores. These techniques suffer from the same
parameter dependency, as was shown by \citet{SchneiderBrooks2004} and
more recently by \citet{PinedaRosolowsky2009}.

\subsection{Numerical resolution study for purely thermal support}
\label{sc:num_res}

\begin{figure*}[t]
\subfigure[solenoidal forcing, $N_{\mathrm{BE}}=1.2\times10^{5}$, thermal support]{\label{fg:resSOL}\includegraphics[width=0.49\textwidth]{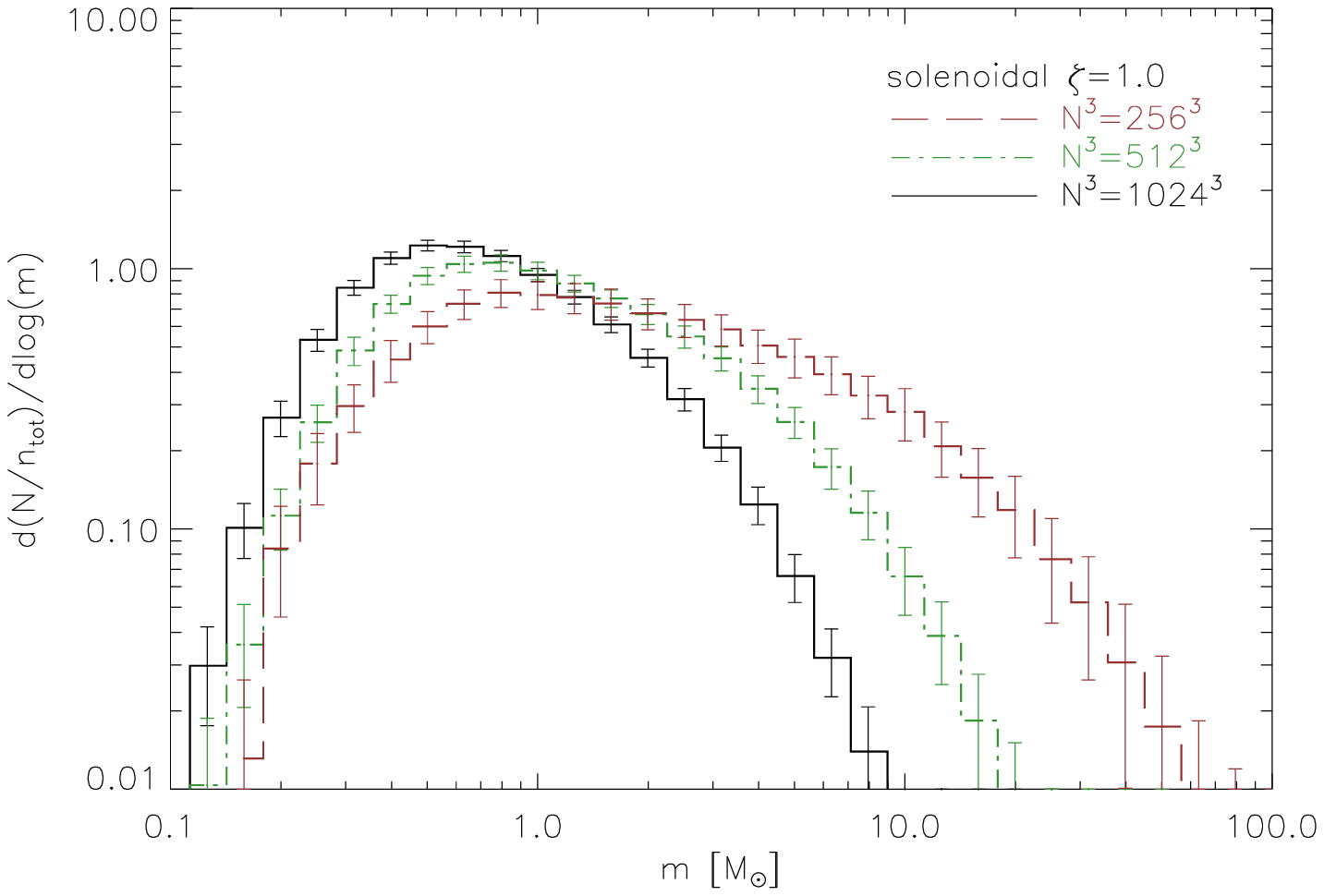}}\hfill
\subfigure[compressive forcing, $N_{\mathrm{BE}}=1.2\times10^{5}$, thermal support]{\label{fg:resDIL}\includegraphics[width=0.49\textwidth]{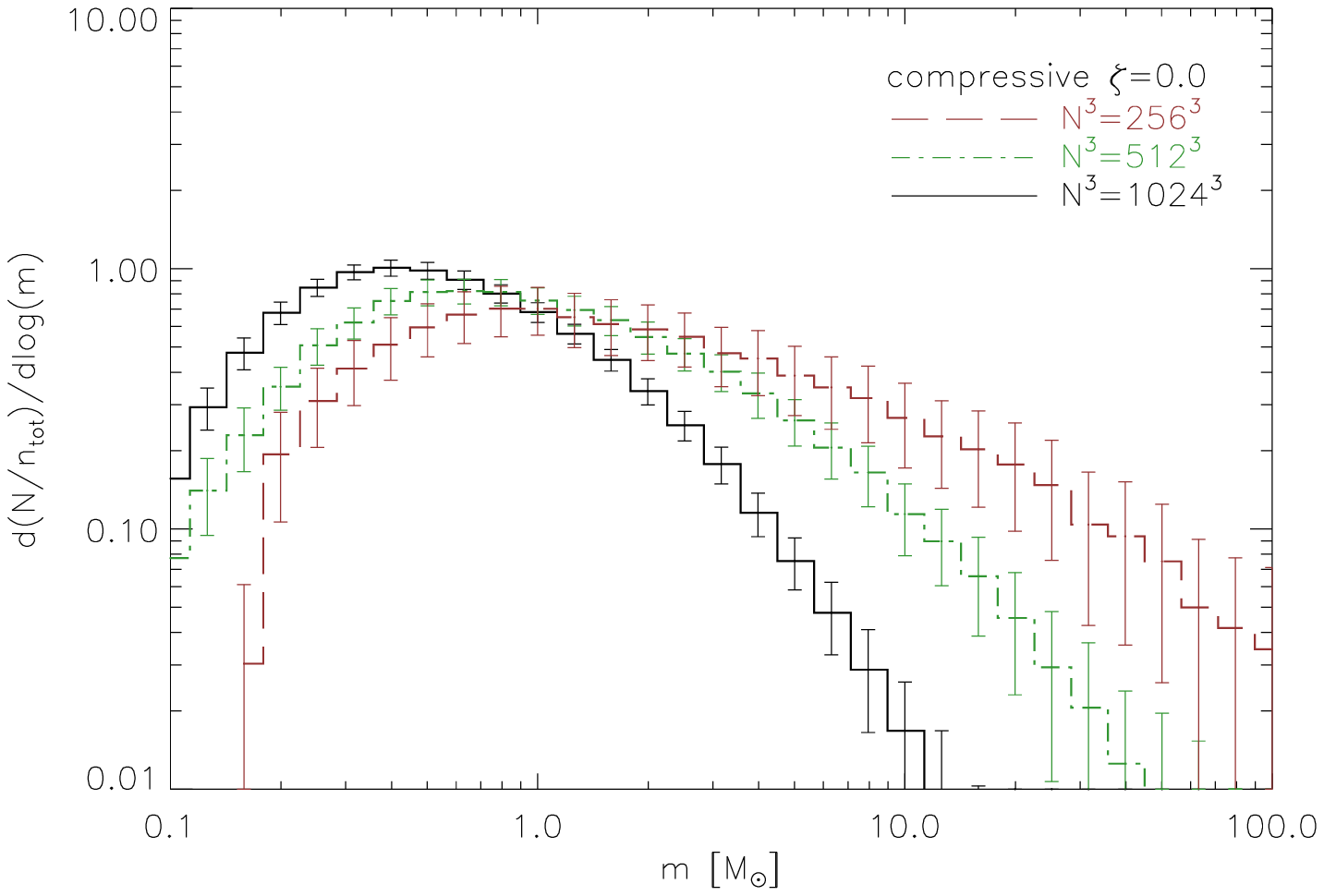}}
\subfigure[solenoidal forcing, $N_{\mathrm{BE}}=10^{3}$, thermal support]{\includegraphics[width=0.49\textwidth]{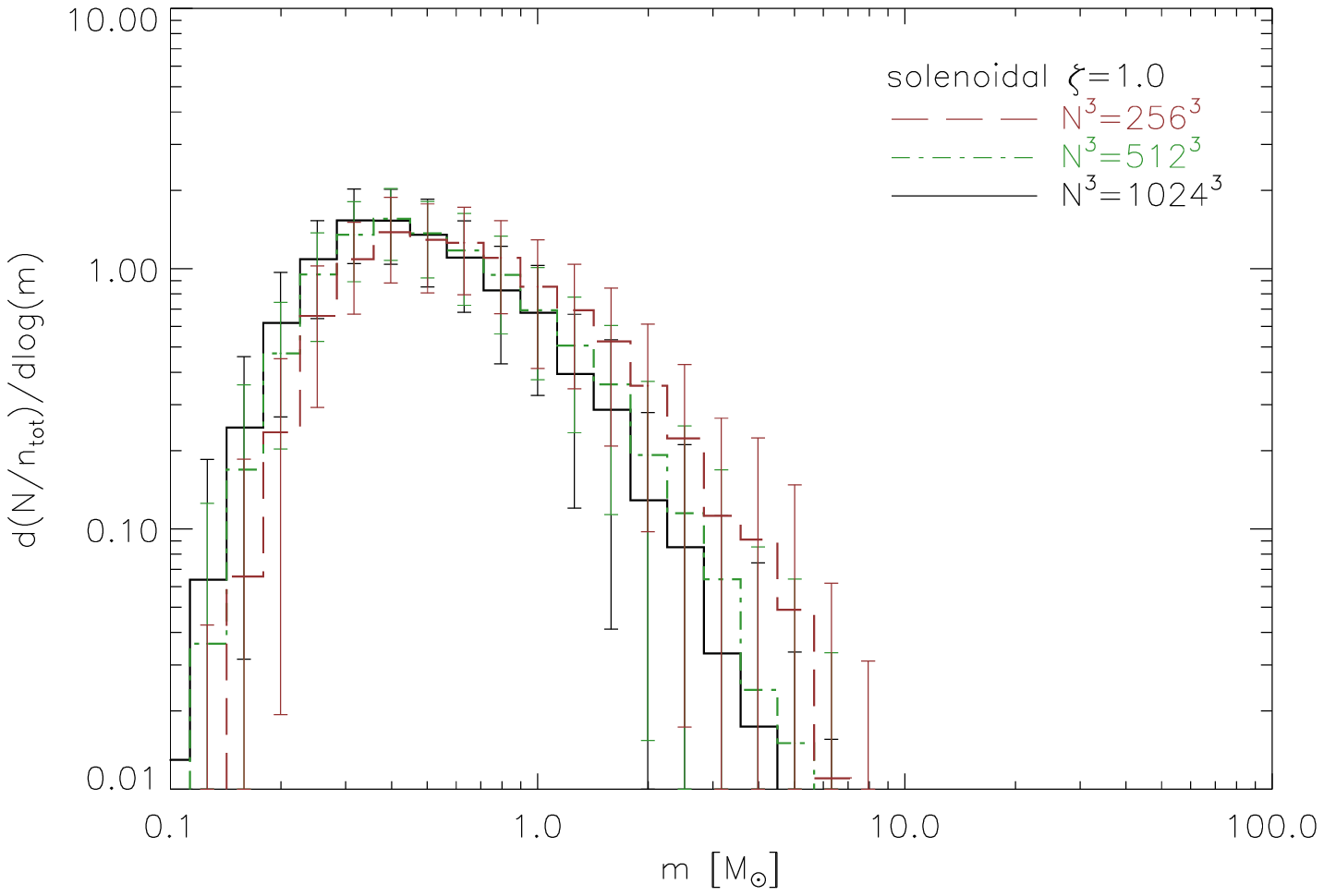}}\hfill
\subfigure[compressive forcing, $N_{\mathrm{BE}}=10^{3}$, thermal support]{\includegraphics[width=0.49\textwidth]{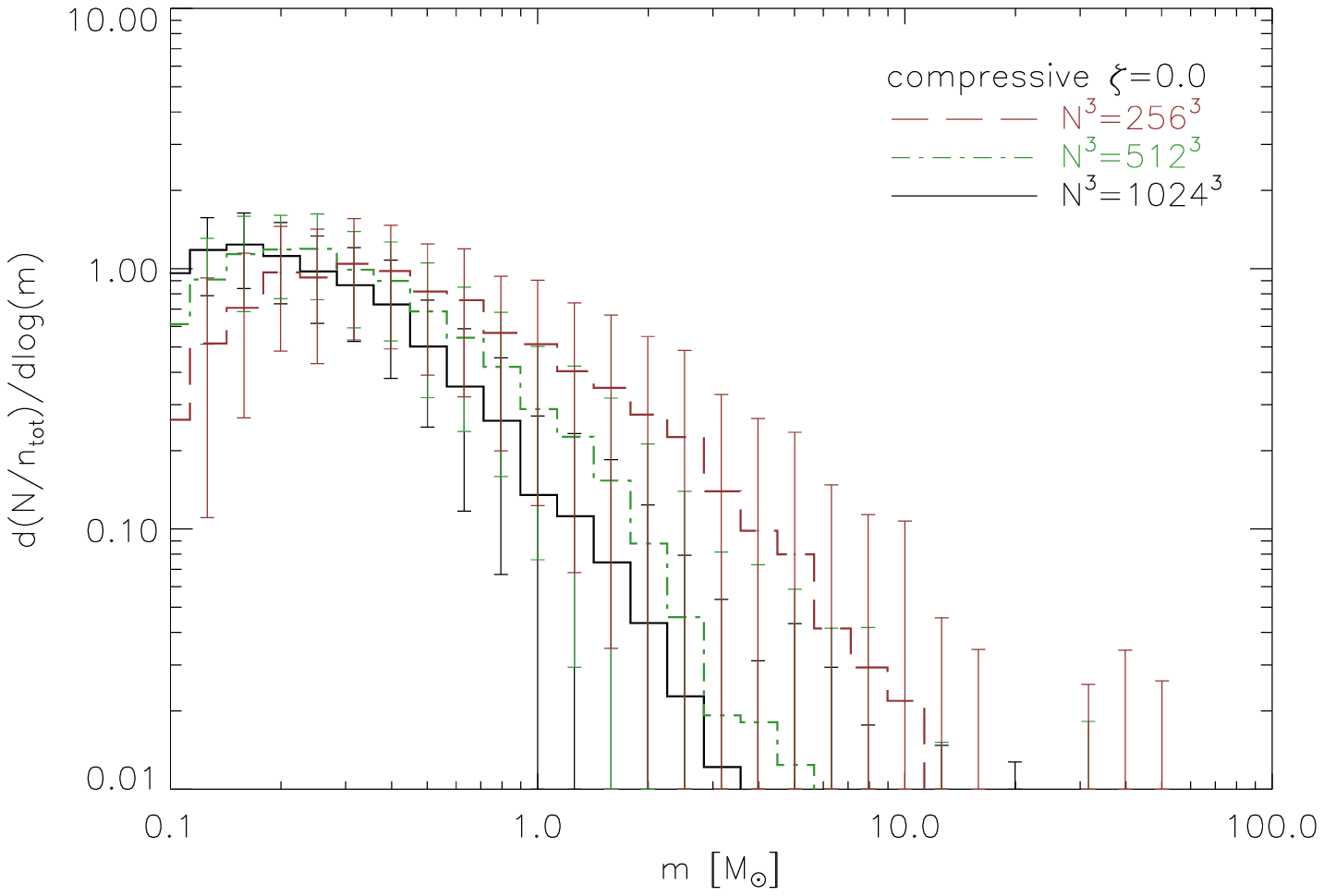}}
\caption{Core mass distributions for solenoidal (left) and compressive forcing (right) at numerical resolutions of $256^3$ (dashed), $512^3$ (dot--dashed) and $1024^3$ (solid), normalised to the total number of cores, $N_{\mathrm{tot}}$. Error bars indicate 1$\sigma$ temporal fluctuations.}
  \label{fg:resolution}
\end{figure*}

Fig. \ref{fg:resolution} indicates that the mass distributions depend very sensitively on the grid resolution of the simulation. For $N_{\mathrm{BE}}=1.2\times10^{5}$,  the high-mass tails steepen systematically with increasing grid resolution for both solenoidal and compressive forcing, showing no sign of convergence in the sense of an asymptotic limit, even for the highest resolution of $1024^{3}$ grid points. The mass distributions for $N_{\mathrm{BE}}=10^{3}$
appear to be better converged, at least in the case of solenoidal forcing. The statistical properties of the core mass distributions are summarised in Table \ref{table:summary}. The mean mass $\overline{m}$ shifts to lower values with increasing resolution, approaching a time averaged value of $\left< \overline{m}\right> \approx 1\ \mathrm{M_{\sun}}$ for the $1024^3$ runs, while the width $\sigma_{m}$ of the distributions decreases. This is caused by the fragmentation of the cores found in the low resolution simulations into smaller and denser cores in the high resolution simulations which also satisfy the condition for gravitational instability. The sum of all core masses, $m_{\mathrm{cores}}$, is a few percent of the total box mass $m_{\mathrm{tot}}$. 

\begin{table*}[th]
\caption{Time averaged properties of the mass distributions for solenoidal and compressive forcing calculated with the physical scalings applied in PN07. 
}
\label{table:summary}      
\centering                         
\begin{tabular}{ccccccccc}        
\hline                
\multicolumn{9}{c}{solenoidal forcing, $\zeta=1.0$}\\
\hline
grid points	 & $\left<\overline{m}\right>\ [\mathrm{M_{\sun}}]$ & $\left<\sigma_{m}\right>\ [\mathrm{M_{\sun}}]$ 	 	& $\left<n_{\mathrm{tot}}\right>$ & $\left<m_{\mathrm{cores}}/ m_{\mathrm{tot}}\right>\ [\%]$ & $x_{\mathrm{LS}}$  & $\left<x_{\mathrm{MLE}}\right>$ & $\left<m_{\mathrm{min}}\right>\ [\mathrm{M_{\sun}}]$ & $\left<p\right>$\\  
\hline\hline                        
   $\mathrm{256}^3$ 	& $4.04\pm0.18$ & $7.00\pm0.69$	  & $1540\pm57$ & $5.1\pm0.2$&$2.1\pm0.5$ &  $1.8\pm0.5$ & $9.8\pm4.9$&  $0.03\pm0.07$	\\      
   $\mathrm{512}^3$ 	& $1.83\pm0.04$	& $2.26\pm0.10$   &$3918\pm103$ &$5.9\pm0.2$ & $3.1\pm0.3$	 & $2.7\pm0.7$ & $6.1\pm2.6$& $0.06\pm0.09$		\\
   $\mathrm{1024}^3$	& $1.08\pm0.02$	& $1.05\pm0.04$   &$7082\pm219$ & $6.3\pm0.2$&$3.1 \pm0.2$  &$3.2\pm0.7$ & $3.5\pm0.9$&  $0.21\pm0.26$\\ 
\hline                             
\multicolumn{9}{c}{compressive forcing, $\zeta=0.0$}\\
\hline
   $\mathrm{256}^3$ 	& $9.7\pm8.0$ 	& $34.36\pm29.67$&$590\pm57$ & $4.0\pm0.7$&$1.0\pm0.3$ &$1.1\pm0.3$	& $9.7\pm8.0$	& $0.05\pm0.07$	\\
   $\mathrm{512}^3$ 	& $2.41\pm0.19$	& $5.54\pm1.48$	&$1859\pm138$ & $3.7\pm0.3$&$1.6\pm0.2$ & $1.6\pm0.3$ & $5.2\pm2.7$ & $0.18\pm0.22$	\\
   $\mathrm{1024}^3$	& $0.98\pm0.04$ & $1.44\pm0.21$	& $4985\pm429$& $4.0\pm0.3$&$2.1\pm0.2$ & $2.2\pm0.3$	& $2.7\pm0.8$	& $0.23\pm0.21$ \\ 
\hline                                   
\end{tabular}
\end{table*}
An important property of the  mass distributions is the slope $x$ of the high-mass wings of
$\dd N/\dd\log m$, which is related to the exponent $\alpha$ of the linear distribution via $x=\alpha-1$.
Assuming that there is a power law range, we applied error-weighted least squares fits to the high-mass power-law regime ($256^3$: $m \geq 15\mathrm{M_{\sun}}$, $512^3$: $m \geq 8\mathrm{M_{\sun}}$, $1024^3$: $m \geq 3\mathrm{M_{\sun}}$) in Fig. \ref{fg:resolution}. The power-law exponents obtained by this method are referred to as $x_{\mathrm{LS}}$.
For the highest resolution we find $x_{\mathrm{LS}}=3.1\pm 0.2$ for solenoidal forcing and a shallower slope of $x_{\mathrm{LS}}=2.1\pm 0.2$ for compressive forcing. The mass distributions of the solenoidal runs tend to show a higher degree of convergence. The differences in the slopes between the three solenoidal simulations decrease with increasing resolution. The slope $x_{\mathrm{LS}}$ obtained from the $512^3$ simulation is steeper by a value of $1.0$ than the $256^3$ while the values of $x_{\mathrm{LS}}$ for the resolutions of $512^3$ and $1024^3$ are nearly identical in the solenoidal case. In contrast, the slopes of the mass distributions obtained from the pure compressive forcing increase by a value of $0.6$ ($256^3\rightarrow 512^3$) and by a value of $0.5$ ($512^3\rightarrow 1024^3$) indicating slower convergence than in the solenoidal forcing case.

We also used the method of maximum likelihood estimation (MLE) (for details see Appendix \ref{ap:mlemethod}) inspired by the work of \citet{Clauset2007} to determine the slope $x_{\mathrm{MLE}}$ of the core mass distributions and to check if the distribution really follows a power law. 
This method avoids biases that can result from least-square fits \citep[e.~g.][]{Bauke2007}
and provides an estimate of the mass value $m_{\mathrm{min}}$ above which the power law assumption is made. Furthermore, the probability $p$ that the power-law assumption holds can be calculated. Following \citet{Clauset2007}, we rule out power laws if $p \leq 0.1$. The results of the MLE method applied to the data of our core masses are summarised in Table \ref{table:summary}. For the highest resolution we find a power law slope of $x_{\mathrm{MLE}}=2.2\pm0.3$ for $m\geq m_{\mathrm{min}}=2.7\pm 0.8\ \mathrm{M_{\sun}}$ in the case of compressive forcing
and $x_{\mathrm{MLE}}=3.2\pm0.7$ for $m\geq m_{\mathrm{min}}=3.5\pm 0.9\ \mathrm{M_{\sun}}$ in the solenoidal forcing case. The errors are the 1$\sigma$ temporal fluctuations of these values. The slopes are consistent with the values obtained via least squares fits. At numerical resolutions of $256^3$ and $512^3$, the time averaged p-values are less than $0.1$ for both solenoidal and compressive forcing. Thus, they are not consistent with power laws. However, at a numerical resolution of $1024^3$, we find $p=0.21\pm0.26$ for solenoidal and $p=0.23\pm0.21$ for compressive forcing. Note that the standard deviations of these p-values are very large, such that instantaneous CMFs can either be consistent with power laws or do not exhibit power-law behaviour at the high-mass end of the distribution. We emphasise that it is absolutely necessary to compute time-averaged quantities in order to obtain statistically meaningful results.
Note that the $p$-value can only rule out the power law hypothesis while other distributions could be a better fit even for $p\geq0.1$. The possible inconsistency of the high-mass CMFs with a power law model was also discussed by \citet{BallesterosParedesEtAl2006}.

For $N_{\mathrm{BE}}=10^{3}$, the scatter around the tails of the mass distributions is too large for the MLE method to be applicable. Therefore, we only list the power-law exponents $x_{\mathrm{LS}}$ of the high-mass tails of the CMFs in Table \ref{table:slopes}. Although being quite inaccurate, especially due to the low number statistics, this method can be used to describe the general trend of the CMFs. For the solenoidal simulation with turbulent support and $N_{\mathrm{BE}}=10^{3}$ we do not give a value for the power-law exponent because the error bars are simply too large. 
Decreasing the total number of Bonner-Ebert masses in the box from $N_{\mathrm{BE}}$ from $1.2\times10^{5}$ to $10^{3}$ leaves the fitted value of $x_{\mathrm{LS}}$ relatively unchanged (taking the large error bars into account). The effect of adding turbulent support on the slope of the CMF will be discussed in Section~\ref{sc:turbpressure_massscale}.

\begin{table}[t]
\caption{Least-square estimates of the power-law exponent $x_{\mathrm{LS}}$
obtained from the $1024^{3}$ simulations for different values of $N_{\mathrm{BE}}$ with/without turbulent
pressure included in the core stability criterion.}
\label{table:slopes}      
\centering                         
\begin{tabular}{lccc}        
\hline                
\multicolumn{4}{c}{solenoidal forcing, $\zeta=1.0$}\\
\hline
$\lambda_{\mathrm{J}}^{0}/L$ & $N_{\mathrm{BE}}$ & 
turbulent support & $x_{\mathrm{LS}}$  \\  
\hline\hline                        
0.04 & $1.2\times 10^{5}$ 	& no  	& $3.1 \pm0.2$ \\ 
0.04 & $1.2\times 10^{5}$ 	& yes 	& $2.8 \pm0.2$ \\ 
0.2 & $1.0\times 10^{3}$ 	& no 	& $2.5 \pm0.7$ \\
0.2 & $1.0\times 10^{3}$ 	& yes 	& -        \\ 
\hline                             
\multicolumn{4}{c}{compressive forcing, $\zeta=0.0$}\\
\hline
0.04 & $1.2\times 10^{5}$ 	& no 	& $2.1 \pm0.2$  \\ 
0.04 & $1.2\times 10^{5}$ 	& yes 	& $2.1 \pm0.2$  \\ 
0.2 & $1.0\times 10^{3}$ 	& no 	& $2.0 \pm0.4$ \\ 
0.2 & $1.0\times 10^{3}$ 	& yes 	& $1.2 \pm0.3$ \\ 
\hline                                   
\end{tabular}
\end{table}

\begin{figure*}[ht]
\subfigure[solenoidal forcing, $N_{\mathrm{BE}}=1.2\times10^{5}$, thermal support]{\label{fg:sol_cells}\includegraphics[width=0.49\linewidth]{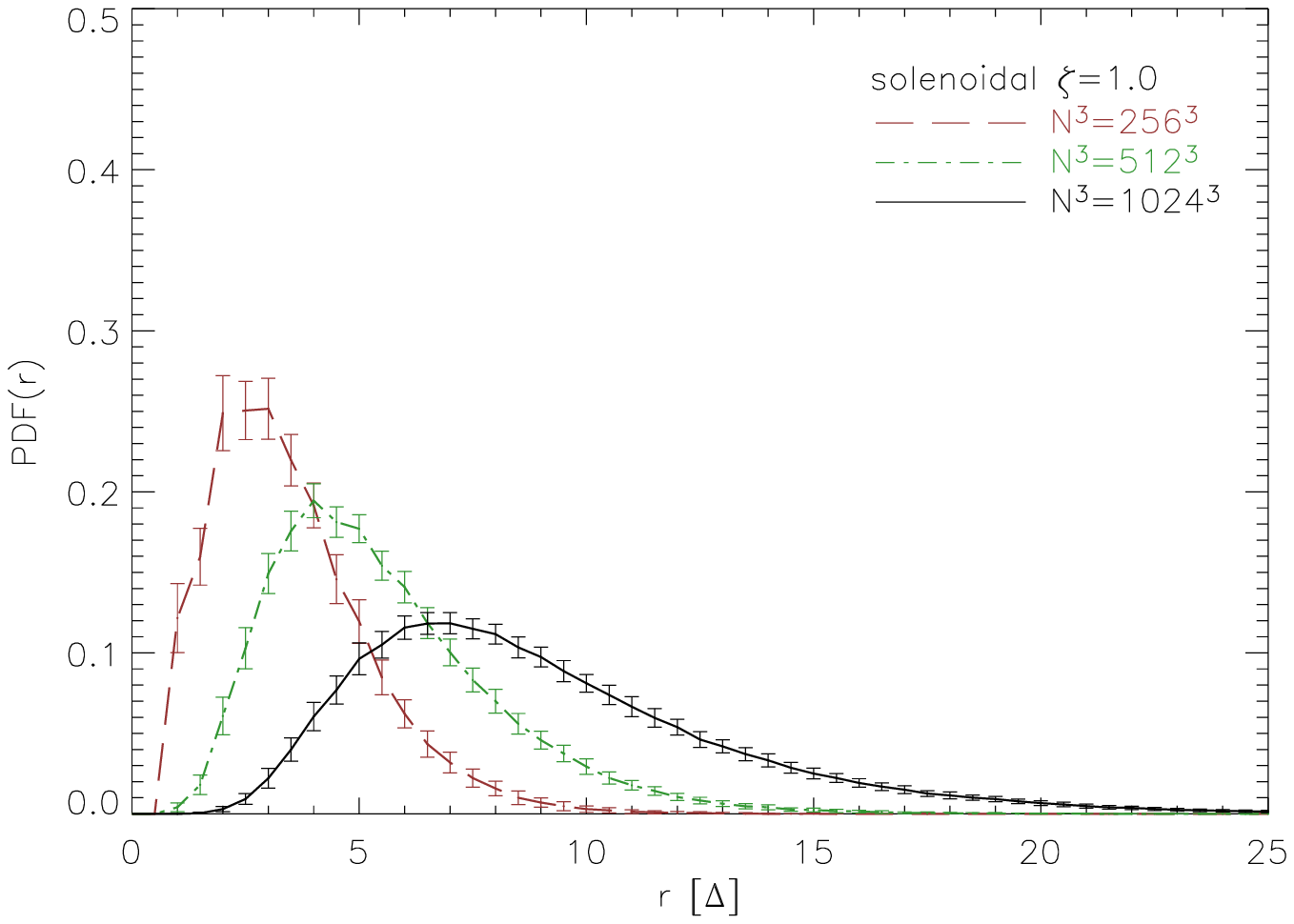}}\hfill
\subfigure[compressive forcing, $N_{\mathrm{BE}}=1.2\times10^{5}$, thermal support]{\label{fg:dil_cells}\includegraphics[width=0.49\linewidth]{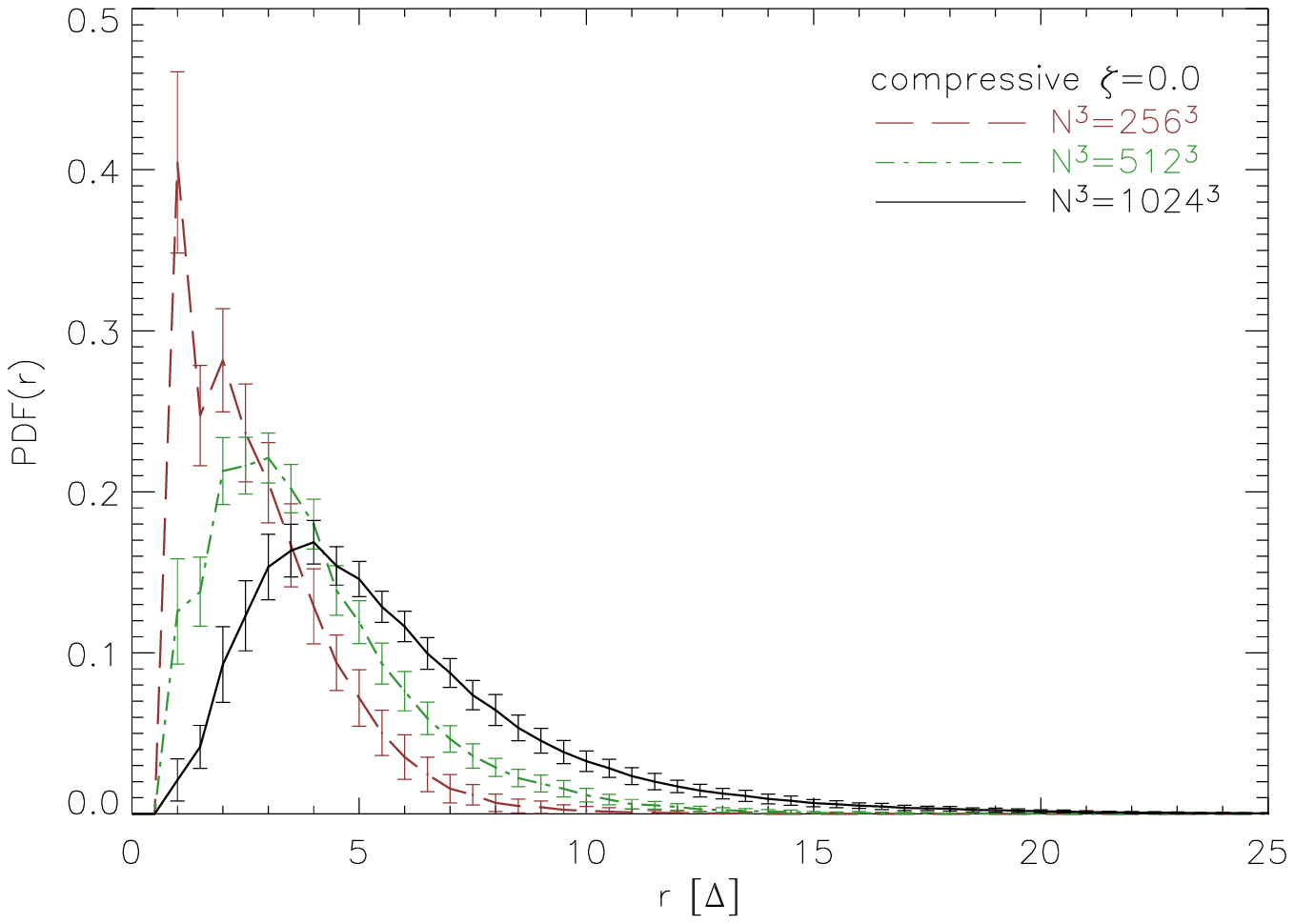}}
\caption{PDF of core size $r$ defined in eqn.~(\ref{eq:size}) for solenoidal \subref{fg:sol_cells} and compressive \subref{fg:dil_cells} forcing for a numerical resolution of $256^3$ (dashed), $512^3$ (dot--dashed) and $1024^3$ (solid) and PN07 scaling. Error bars indicate 1$\sigma$ temporal fluctuations of the PDFs.}
  \label{fg:resolution_sizes}
\end{figure*}

In order to further investigate the effects of numerical resolution on the core properties, we look at the size $r$ of the cores. We define the approximate size of a core as 
\begin{equation}\label{eq:size}
 r=n^{1/3}_{\mathrm{cells}}\Delta,
\end{equation}
where $n_{\mathrm{cells}}$ is the number of grid cells contained in a core and $\Delta=X/N$ with $N=256, 512, 1024$ depending on the numerical grid resolution. We will first consider the
mass scale chosen by PN07, i.~e., $N_{\mathrm{BE}}\approx 1.2\times10^{5}$. Then we will comment on the changes arising from a lower number of Jeans masses in the box (corresponding to $N_{\mathrm{BE}}\approx \times10^{3}$). The typical size of large cores  should be $\sim \lambda_{\mathrm{J}}^{0}$, which is about $5.2\Delta$ for $N=256$ and $20.8\Delta$ for $N=1024$. The fragmentation properties of the gas on length scales less than about $10\Delta$, which encompasses most of the range of core sizes, are affected by numerical smoothing. Consequently, the probability density functions (PDFs) of $r$, which are plotted in Fig. \ref{fg:resolution_sizes}, change substantially with resolution. In particular, one can also see that
the fractions of cores with $r\gtrsim\lambda_{\mathrm{J}}^{0}$ are relatively small, especially for the high-resolution data. At lower resolution, however, these fractions increase, i.~e., more big cores are found relative to the number of smaller cores. Thus, the maxima of the mass distributions
shown in Figure~\ref{fg:resolution} are shifted towards higher masses with decreasing resolution. In addition, the discrepancies in the high-mass wings might be caused by a bias of the clump-finding algorithm to select cores that contain several gravitationally unstable cores.

\begin{table*}[ht]
\caption{Time averaged properties of the core size PDFs for solenoidal and compressive forcing.
}
\label{table:sizes_summary}      
\centering                         
\begin{tabular}{cccccc}        
\hline                
\multicolumn{6}{c}{solenoidal forcing, $\zeta=1.0$}\\
\hline
grid points	 	& $N_{\mathrm{BE}}$ & turbulent support& $\left<\overline{r}\right>\ [\mathrm{cells}]$ & $\left<\sigma_{r}\right>\ [\mathrm{cells}]$ 	 	& $\left<f_{r\leq 2}\right> [\%]$  \\  
\hline\hline                        
   $\mathrm{256}^3$ 	&$1.2\times 10^{5}$ & no  & $3.72\pm0.08\phantom{0}$	&  $1.74\pm0.05\phantom{0}$	& $15.8\pm1.6\phantom{0}$ \\      
   $\mathrm{256}^3$ 	&$1.2\times 10^{5}$ & yes & $3.70\pm0.08$	&  $1.77\pm0.05$	& $16.7\pm1.7$ \\      
   $\mathrm{256}^3$ 	&$1.0\times 10^{3}$ & no  & $10.5\pm0.5$	&  $3.4\pm0.3$		& $0$ 	\\   
   $\mathrm{256}^3$ 	&$1.0\times 10^{3}$ & yes & $12.8\pm1.0$	&  $4.8\pm0.7$		& $0$ 	\\   
   $\mathrm{512}^3$ 	&$1.2\times 10^{5}$ & no  &$5.78\pm0.12$	&  $2.51\pm0.05$	& $1.4\pm0.5$	      \\
   $\mathrm{512}^3$ 	&$1.2\times 10^{5}$ & yes  &$5.82\pm0.13$	&  $2.60\pm0.06$	& $1.6\pm0.5$	      \\
   $\mathrm{512}^3$ 	&$1.0\times 10^{3}$ & no  & $18.4\pm0.8$	&  $6.1\pm0.5$		& $0$	\\
   $\mathrm{512}^3$ 	&$1.0\times 10^{3}$ & yes & $26.3\pm3.0$	&  $10.7\pm1.9$		& $0$	\\
   $\mathrm{1024}^3$	&$1.2\times 10^{5}$ & no  &$9.37\pm0.22$	&  $4.17\pm0.10$	& $0.037\pm0.044$ \\ 
   $\mathrm{1024}^3$	&$1.2\times 10^{5}$ & yes &$9.82\pm0.24$	&  $4.56\pm0.12$	& $0.041\pm0.050$ \\ 
   $\mathrm{1024}^3$	&$1.0\times 10^{3}$ & no  & $34.5\pm1.8$	&  $11.7\pm1.1$		& $0$	\\ 
   $\mathrm{1024}^3$	&$1.0\times 10^{3}$ & yes  & $60.8\pm7.6$	&  $23.1\pm4.3$		& $0$	\\ 
\hline                             
\multicolumn{6}{c}{compressive forcing, $\zeta=0.0$}\\
\hline
   $\mathrm{256}^3$ 	&$1.2\times 10^{5}$ & no & $2.98\pm0.15\phantom{0}$	&  $1.67\pm0.11\phantom{0}$ 	& $34.8\pm3.6\phantom{0}$ 	\\
   $\mathrm{256}^3$ 	&$1.2\times 10^{5}$ & yes & $2.94\pm0.15$	&  $1.68\pm0.11$ 	& $35.9\pm3.6$ 	\\
   $\mathrm{256}^3$ 	&$1.0\times 10^{3}$ & no & $6.0\pm0.6$		&  $3.3\pm0.5$ 		& $2.9\pm2.9$\\
   $\mathrm{256}^3$ 	&$1.0\times 10^{3}$ &yes & $5.8\pm0.6$		&  $3.4\pm0.6$ 		& $3.5\pm3.6$\\
   $\mathrm{512}^3$ 	&$1.2\times 10^{5}$ & no & $4.18\pm0.19$	&  $2.28\pm0.11$	& $14.7\pm3.8$	\\
   $\mathrm{512}^3$ 	&$1.2\times 10^{5}$ & yes & $4.10\pm0.19$	&  $2.29\pm0.12$	& $15.9\pm2.8$	\\
   $\mathrm{512}^3$ 	&$1.0\times 10^{3}$ & no & $9.0\pm0.8$		&  $4.7\pm0.9$		& $0.04\pm0.18$	\\
   $\mathrm{512}^3$ 	&$1.0\times 10^{3}$ & yes & $8.9\pm0.9$		&  $5.0\pm1.2$		& $0.039 \pm 0.2$	\\
   $\mathrm{1024}^3$	&$1.2\times 10^{5}$ & no & $6.07\pm0.37$	&  $3.42\pm0.23$ 	& $3.7\pm1.4$  \\ 
   $\mathrm{1024}^3$	&$1.2\times 10^{5}$ & yes & $5.95\pm0.38$	&  $3.45\pm0.25$ 	& $4.2\pm1.6$  \\ 
   $\mathrm{1024}^3$	&$1.0\times 10^{3}$ & no & $15.1\pm1.9$		&  $9.0\pm1.7$ 		& $0.008\pm0.07$   \\ 
   $\mathrm{1024}^3$	&$1.0\times 10^{3}$ & yes & $16.0\pm2.4$	&  $11.0\pm2.6$		& $0.012\pm0.10$   \\ 
\hline                                   
\end{tabular}
\end{table*}

For a certain variance, $\sigma^{2}$, of the density fluctuations, we expect that cores of size $\sim\sigma^{-1/2}\lambda_{\mathrm{J}}^{0}$ are most frequent, because the Jeans length changes with the inverse square root of the mass density. From the values of $\sigma$ for solenoidal and compressive forcing \citep[see][Table 1]{FederDuv09}, it follows that $(\sigma_\mathrm{comp}/\sigma_\mathrm{soln})^{1/2}=\sqrt{5.9/1.9}\approx1.76$. This is in agreement with the relative peak positions of the distributions for $N=1024$ in Fig.~\ref{fg:resolution_sizes}. Moreover, the mean size $\overline{r}$ of the cores for solenoidal forcing is larger by about the same factor than $\overline{r}$ for compressive forcing, as one can see from the values listed in Table \ref{table:sizes_summary}.

The minimal core size is roughly given by the peak densities $\rho_{\mathrm{max}}$ in the turbulent gas. The definition of the Jeans length implies $r_{\mathrm{min}}\sim(\bar{\rho}/\rho_{\mathrm{max}})^{1/2}\lambda_{\mathrm{J}}^{0}$. Since
$\rho_{\mathrm{max}}$ is roughly $500\bar{\rho}$ for solenoidal forcing and $10^{4}\bar{\rho}$ for compressive forcing \citep{Federrath2009}, it follows that $r_{\mathrm{min}}\sim0.9\Delta$ and $r_{\mathrm{min}}\sim 0.2\Delta$, respectively, for the highest resolution. Consequently,
the smallest cores are marginally resolved in the $1024^{3}$ simulation with solenoidal forcing, while they are definitely below the resolution limit for compressively driven turbulence. As an indicator,
we calculated the fractions $f_{r\leq 2}$ of cores with $r \leq 2$ (see Table~\ref{table:sizes_summary}). Indeed, significant fractions $f_{r\leq 2}$ are obtained for all resolutions in the case of compressive forcing. In the solenoidal simulations, on the other hand, $f_{r\leq 2}$ drops from $15.8\%$ ($256^3$) to $0.037\%$ in the $1024^3$ simulation. These trends are also visible in Fig. \ref{fg:resolution_sizes}. 

\begin{figure*}[ht]
\subfigure[solenoidal forcing, $N_{\mathrm{BE}}=10^{3}$, thermal support]{\label{fg:sol_cells_1e3}\includegraphics[width=0.49\linewidth]{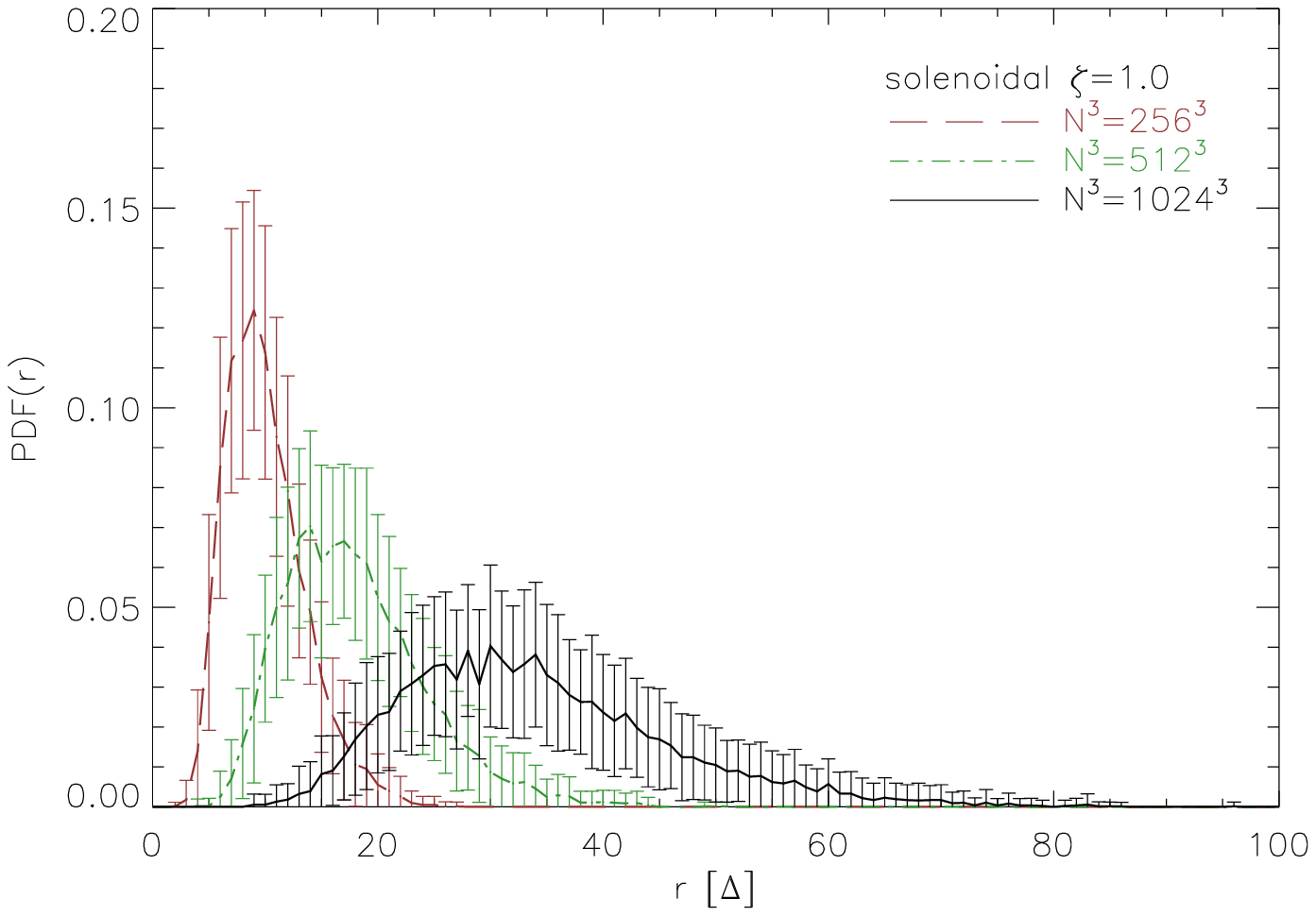}}
\subfigure[compressive forcing, $N_{\mathrm{BE}}=10^{3}$, thermal support]{\label{fg:dil_cells_1e3}\includegraphics[width=0.49\linewidth]{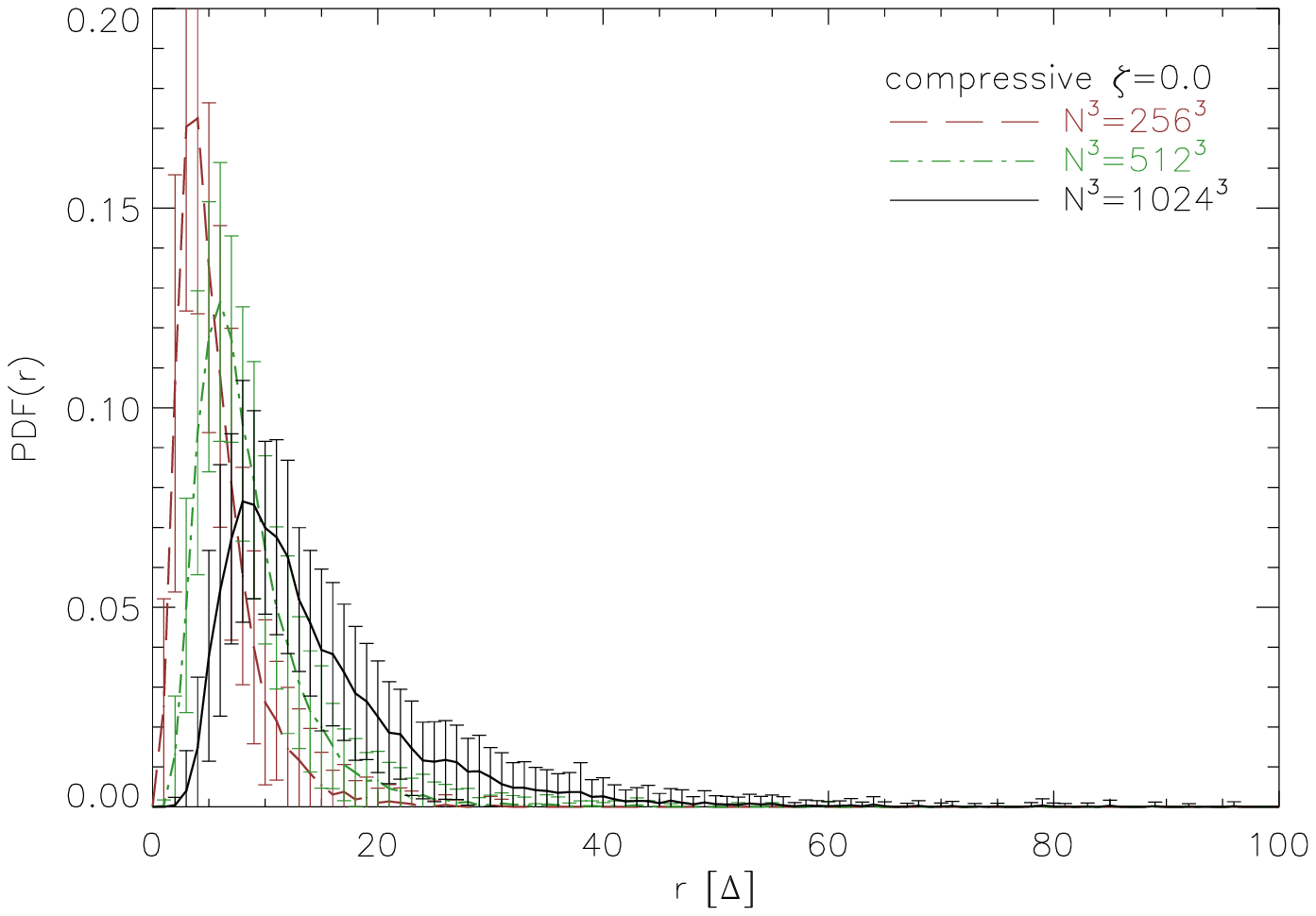}}\hfill
\subfigure[solenoidal forcing, $N_{\mathrm{BE}}=10^{3}$, turbulent support]{\label{fg:sol_cells_1e3_pt}\includegraphics[width=0.49\linewidth]{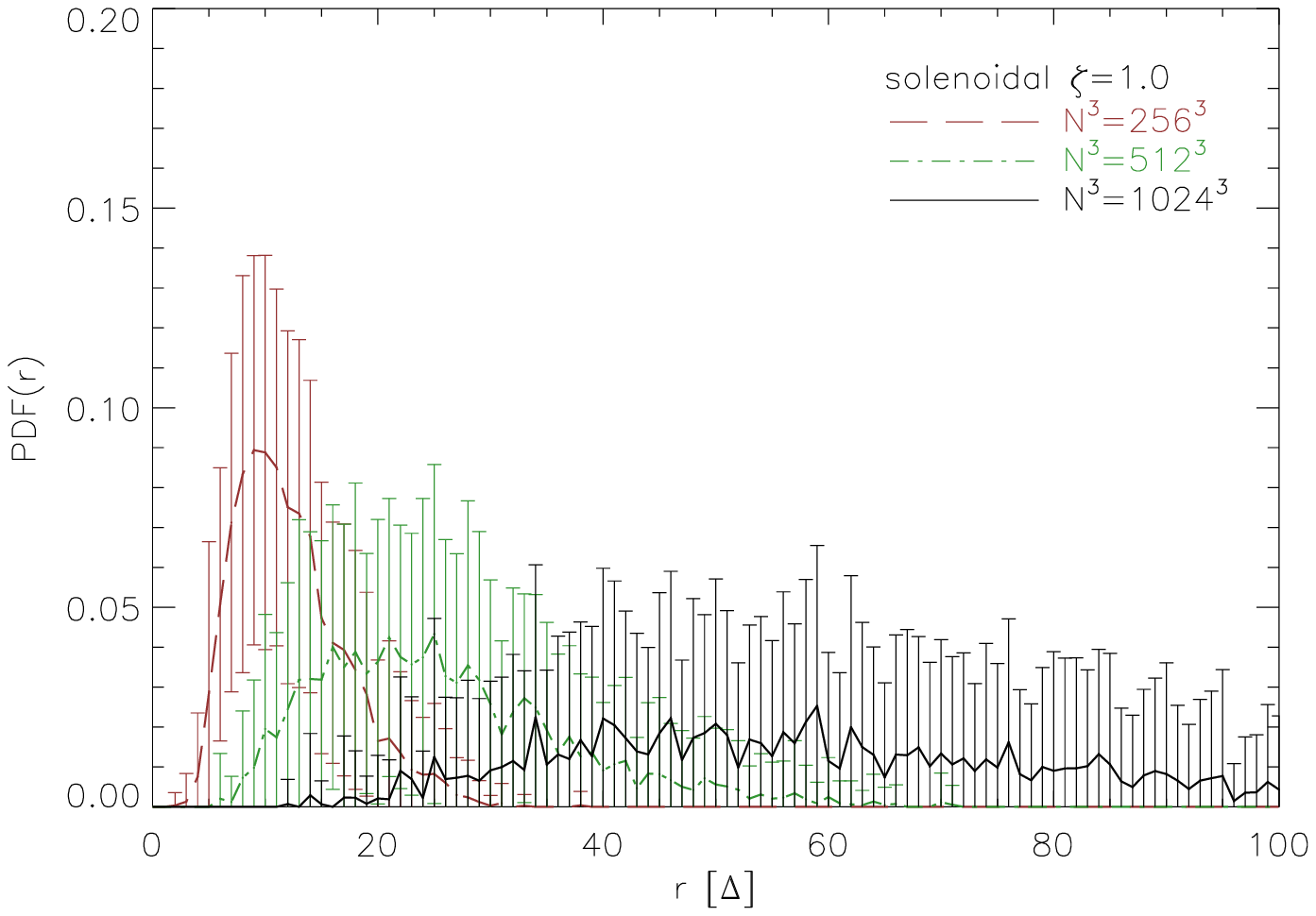}}
\subfigure[compressive forcing, $N_{\mathrm{BE}}=10^{3}$, turbulent support]{\label{fg:dil_cells_1e3_pt}\includegraphics[width=0.49\linewidth]{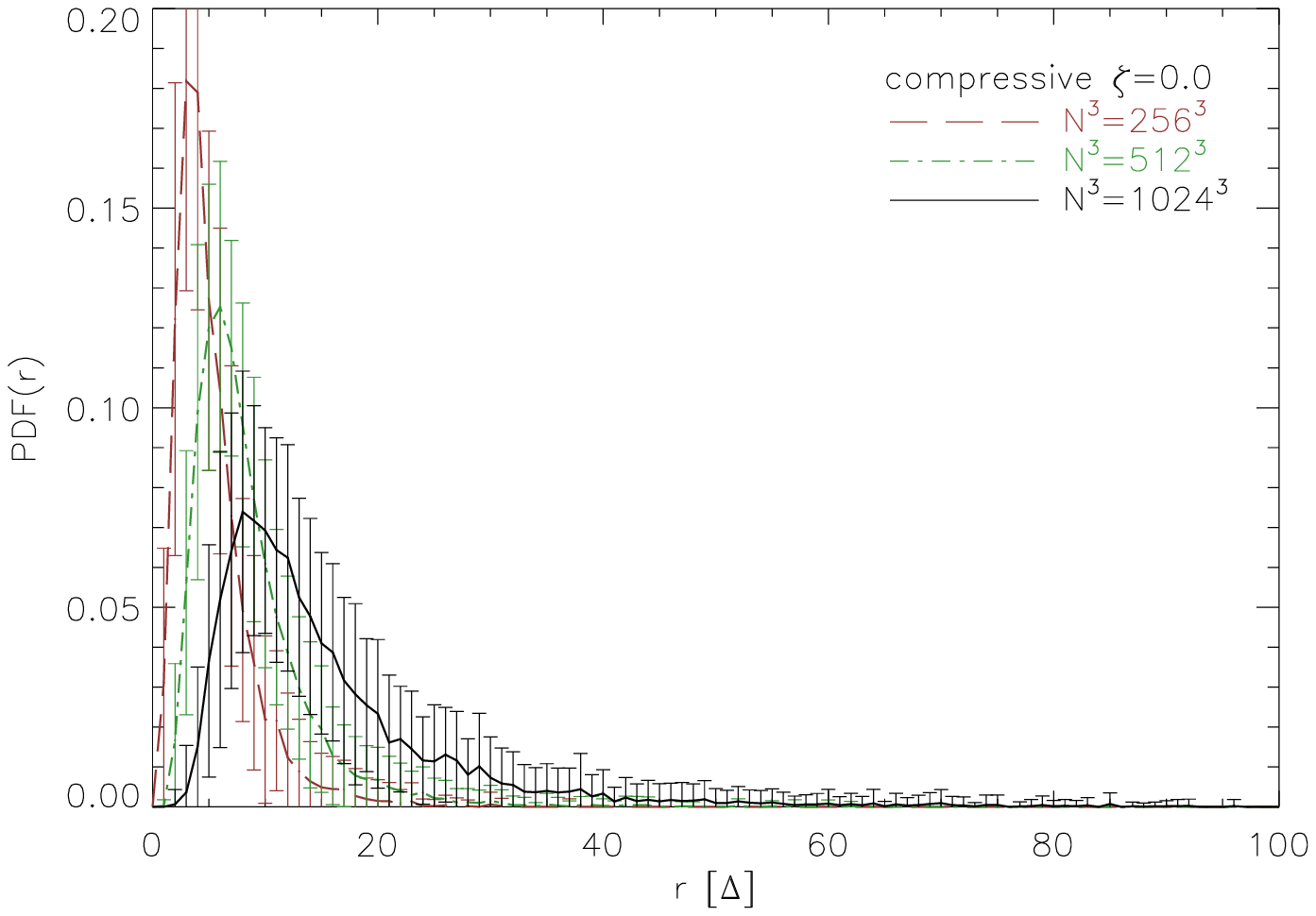}}\hfill
\caption{PDFs of core size distribution as in Fig. \ref{fg:resolution_sizes} but for $N_{\mathrm{BE}}=10^ {3}$ with/without turbulent pressure support
(see Sect.~\ref{sc:hc_theory} and~\ref{sc:scaling}).}
  \label{fg:resolution_sizes2}
\end{figure*}

Setting $N_{\mathrm{BE}}=10^{3}$, the estimate of the minimal core size
from the peak density yields $r_{\mathrm{min}}\approx 4.6\Delta$ for solenoidal forcing and
$1.0\Delta$ for compressive forcing. In the solenodial case, $f_{r\leq 2}$ drops to zero for the
highest resolution. Compared to the PN07 setting, the cores are in general larger
(see Table~\ref{table:sizes_summary} and the top panels of Fig.~\ref{fg:resolution_sizes2}). 
Accordingly, the resolution dependence of the mass distributions is less pronounced, particularly, in the case of solenoidal forcing. However, comparing the mass distributions without normalization
in Fig. \ref{fg:1024dilsol_thermal} and \ref{fg:1024dilsol_1e3_thermal}, one can see that the total number of cores decreases by two orders of magnitude if the mass scale is chosen such that $N_{\mathrm{BE}}=10^{3}$. As in the case $N_{\mathrm{BE}}\approx 1.2\times10^{5}$, we find that the mean core size for solenoidal forcing is roughly twice as large as for compressive forcing.

\subsection{Influence of the turbulent pressure}
\label{sc:turbpressure_massscale}

As explained in Section~\ref{sc:scaling}, we used the effective speed of sound according to eqn.~(\ref{eq:cseff}) in the definition of the Bonnor-Ebert mass, eqn.~(\ref{eq:bonnorebert}), to compute the mass distributions of cores with thermal and turbulent support. For brevity, we will use the
term turbulent support, for which it is understood that the instability criterion is based on the sum of thermal and turbulent pressure.

\begin{figure*}[ht]
\centering
\subfigure[$N_{\mathrm{BE}}=1.2\times10^{5}$, thermal support]{\label{fg:1024dilsol_thermal}\includegraphics[width=0.49\linewidth]{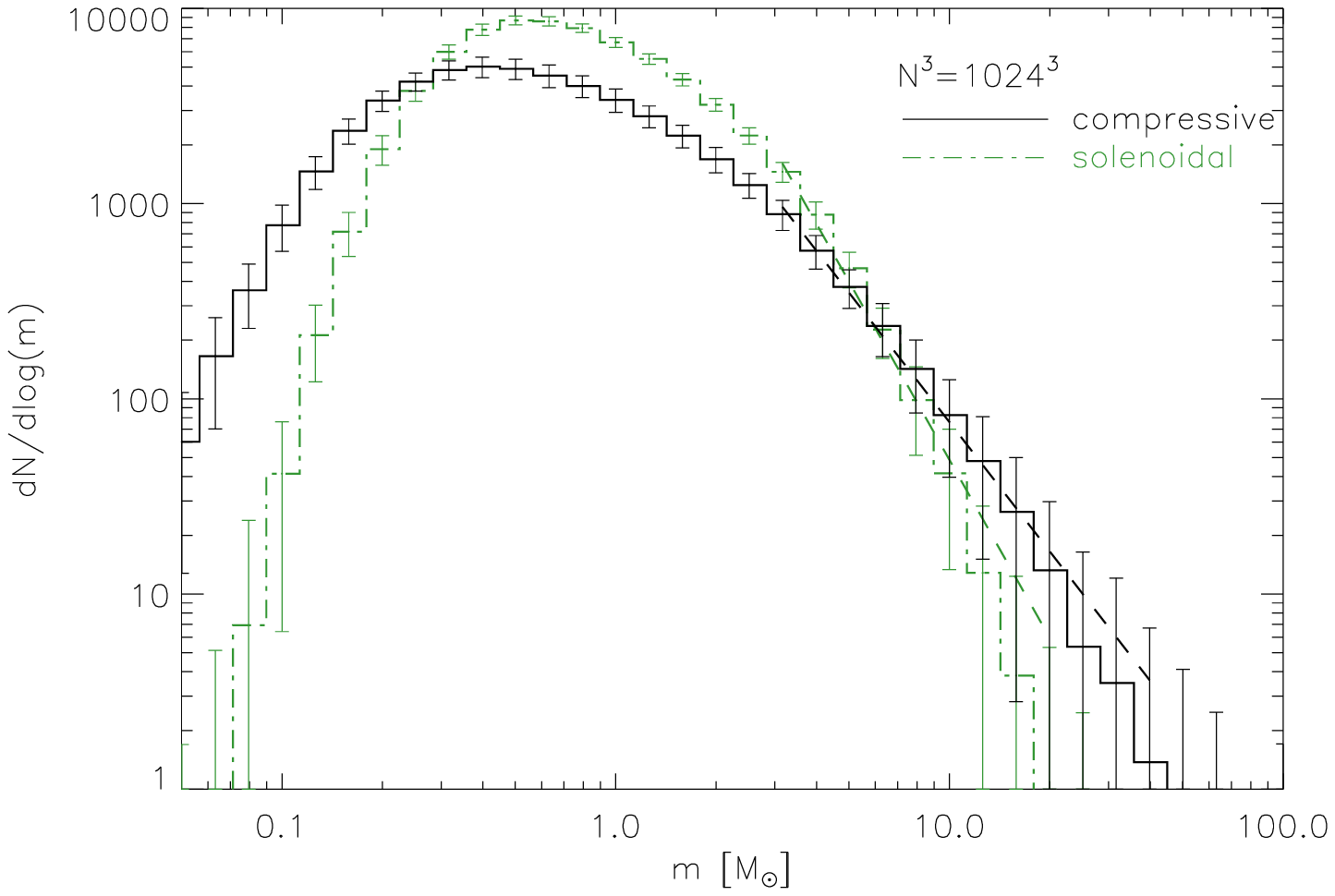}}\hfill
\subfigure[$N_{\mathrm{BE}}=1.2\times10^{5}$, turbulent support]{\label{fg:1024dilsol_turbulent}\includegraphics[width=0.49\linewidth]{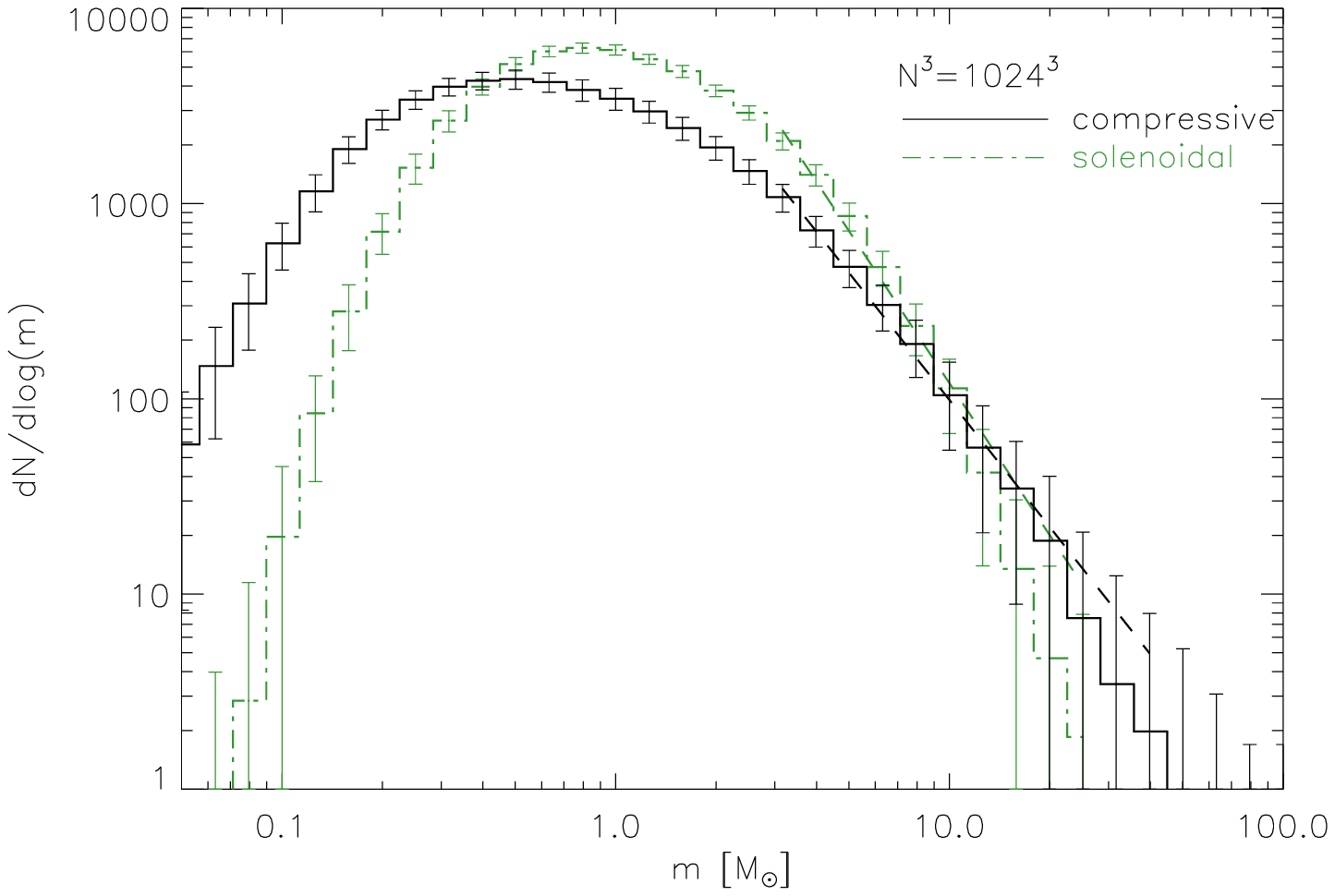}}
\subfigure[$N_{\mathrm{BE}}=10^{3}$, thermal support]{\label{fg:1024dilsol_1e3_thermal}\includegraphics[width=0.49\linewidth]{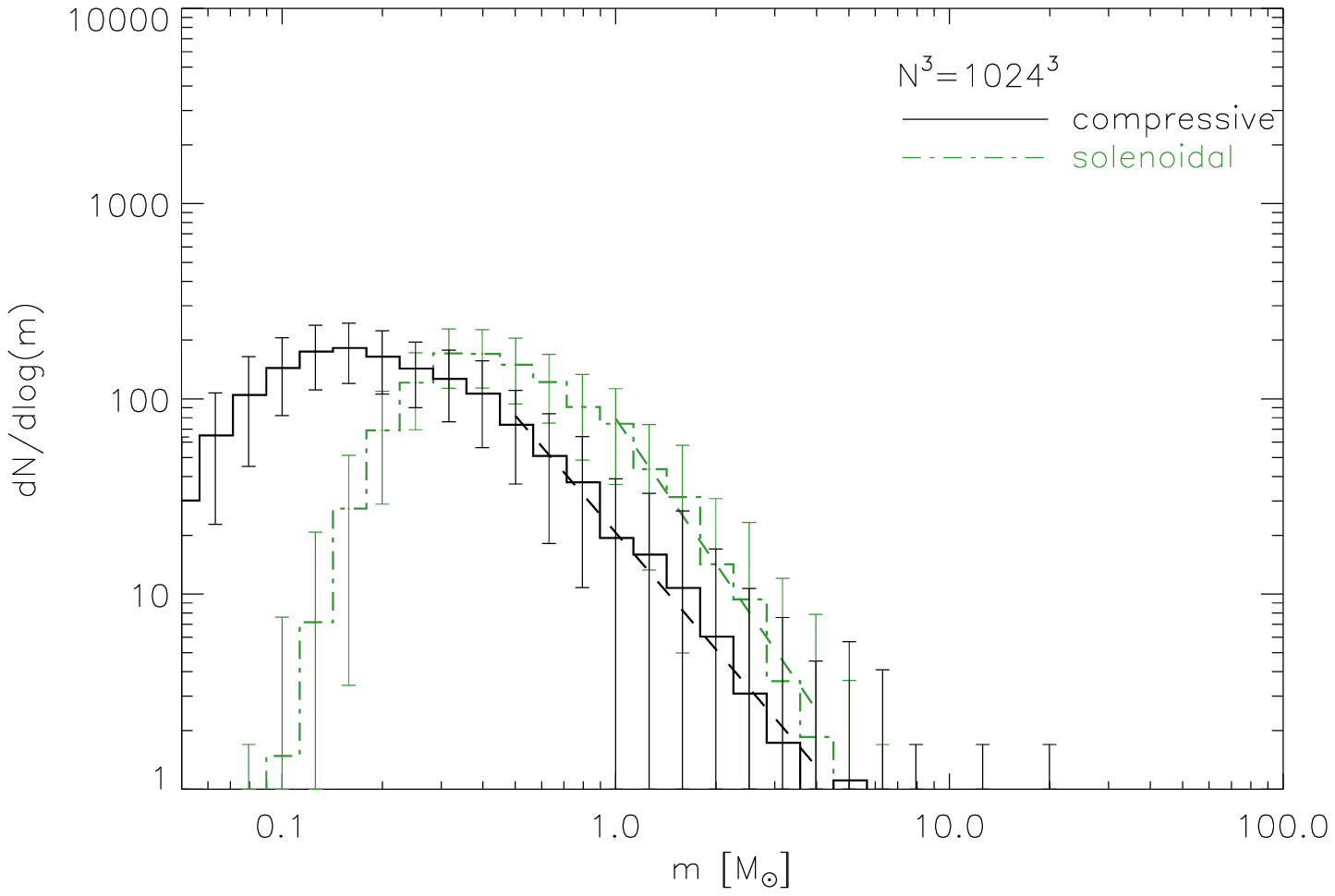}}\hfill
\subfigure[$N_{\mathrm{BE}}=10^{3}$, turbulent support]{\label{fg:1024dilsol_1e3_turbulent}\includegraphics[width=0.49\linewidth]{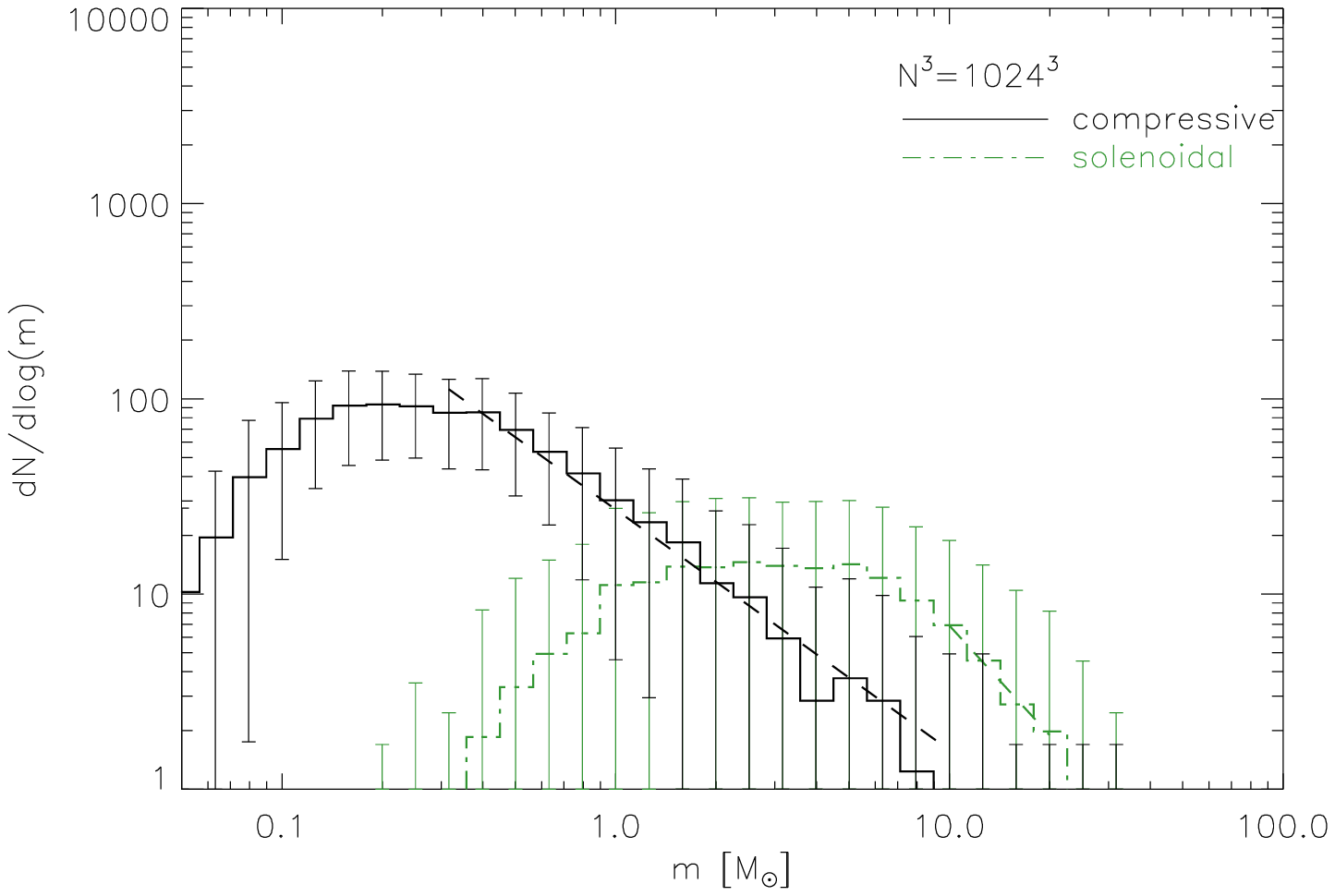}}
\caption{CMFs of compressive (solid) and solenoidal (dot--dashed) forcing for different values of $N_{\mathrm{BE}}$ with/without turbulent support (see Sect.~\ref{sc:hc_theory} and~\ref{sc:scaling}) and a grid resolution of $1024^{3}$. The
least-square fits to the high-mass tails are shown as dashed lines.}
\label{fg:dilsol1024}
\end{figure*} 

Fig. \ref{fg:1024dilsol_turbulent} and \ref{fg:1024dilsol_1e3_turbulent} compares the CMFs for $N_{\mathrm{BE}}=1.2\times10^ {5}$ and $N_{\mathrm{BE}}=10^ {3}$ with turbulent support for the highest grid resolution of $1024^{3}$. While turbulent support has no significant impact on the CMF for $N_{\mathrm{BE}}=1.2\times10^{5}$, it clearly changes the CMF for $N_{\mathrm{BE}}=10^{3}$. For solenoidal forcing, turbulent support leads to a large shift of the peak in the CMF, corresponding to the pronounced change of the core size distribution. As one can see in Fig. \ref{fg:1024dilsol_1e3_turbulent}, however, the temporal fluctuations are larger than the time average.
This is a consequence of the small total number of cores. In the compressive case with turbulent support, the peak of the CMF is relatively robust, but we observe a flattening of the high-mass tail
from $x_{\mathrm{LS}}\approx 2.0$ for the CMF without turbulent support  to $x_{\mathrm{LS}}\approx 1.2$ (see Table~\ref{table:slopes}), which is close to the Salpeter value of $1.35$.

Table \ref{table:sizes_summary} gives a summary of the basic properties of the core size distribution for all possible combinations of grid resolution, turbulent/thermal support and $N_{\mathrm{BE}}$.  
If turbulent support is included, $f_{r\leq 2}$ becomes negligible both for solenoidal and for
compressive forcing if $N=1024$. The core size PDFs are plotted in Fig. \ref{fg:resolution_sizes2}. The large error bars are the result of the much smaller number of cores found in each snapshot compared to $N_{\mathrm{BE}}=1.2\times10^{5}$. Including turbulent support, the probability
to find large cores increases substantially for solenoidal forcing, while the PDF for compressive forcing
does not change appreciably. In the case of solenoidal forcing, the largest effect on the core size is observed for the highest resolution of $1024^{3}$ grid cells. Thus, the support of cores against gravitational collpase is greatly enhanced by the turbulent pressure on length scales that are sufficiently large compared to the grid scale. In Sect.~\ref{sc:comparison_turb}, we show that the turbulent pressure exceeds the thermal pressure on length scales above $30\Delta$, i.~e., outside the range of length scales that are subject to significant numerical dissipation. 

\section{Comparison to theoretical models}
\label{sc:comparison}

In this Section, the core mass distributions computed with the clump-finding algorithm are compared to theoretical predictions of \citet{PadoanNordlund2002} as well as \citet{HennebelleChabrier2008}.
The influence of the width of the density PDF resulting from different forcing was investigated by
\citet{HenneChab09}.
Because \citet{Federrath2008,FederDuv09} showed that the density PDFs are not exactly log-normal, we evaluated equations~(\ref{eq:PN_mass_spect}), (\ref{eq:HC_mass_spect_log_therm}), and~(\ref{eq:HC_mass_spect_log}) for the numerical PDFs calculated from the simulation data. The PDFs
for the three numerical resolutions are plotted in \citet[][Fig.~4 and~6]{FederDuv09}. The resulting
CMFs are thus semi-analytic. This is particularly important for the case of compressively driven turbulence, for which the core mass distributions calculated from log-normal fits to the PDFs of the mass density deviate substantially from the semi-analytic distributions based on the numerical PDFs \citep[see also][]{SchmidtEtAl2009}.
For the comparison with the data from our clump-finding analysis, we normalise the distributions
with the total number of cores $N_{\mathrm{tot}}$, which is calculated from eqn.~(\ref{eq:number_tot}).
Allowing for an arbitrary geometry factor $a_{\mathrm{J}}$ in the definition of the thermal Jeans mass \citep[see][]{HennebelleChabrier2008}, we set $a_{\mathrm{J}}=1.18$ and identify $M_\mathrm{J}^{0}$ with the Bonnor-Ebert mass $m_{\mathrm{BE}}(\rho)M_{\sun}$ (see
eqn.~\ref{eq:bonnorebert}). Thereby, we have a criterion for gravitational instability on the basis of the thermal pressure that is consistent with the clump-finding algorithm. 

\begin{figure*}[t]
\centering
\subfigure[ thermal support, $\lambda_{\mathrm{J}}^{0}/L=0.04$, $N_{\mathrm{BE}}=2\times 10^{5}$]{\includegraphics[width=0.49\linewidth]{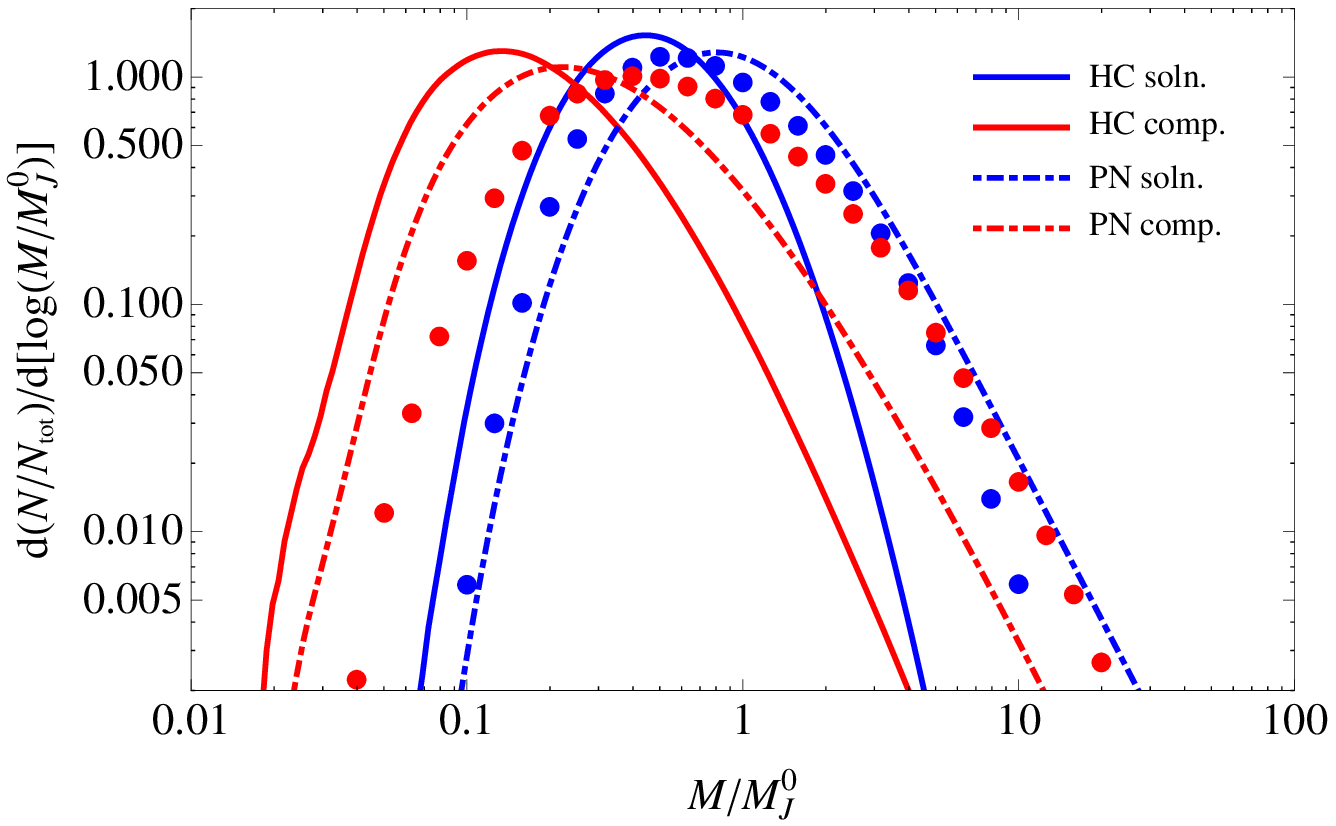}}
\subfigure[ thermal support, $\lambda_{\mathrm{J}}^{0}/L=0.2$, $N_{\mathrm{BE}}=10^{3}$]{\includegraphics[width=0.49\linewidth]{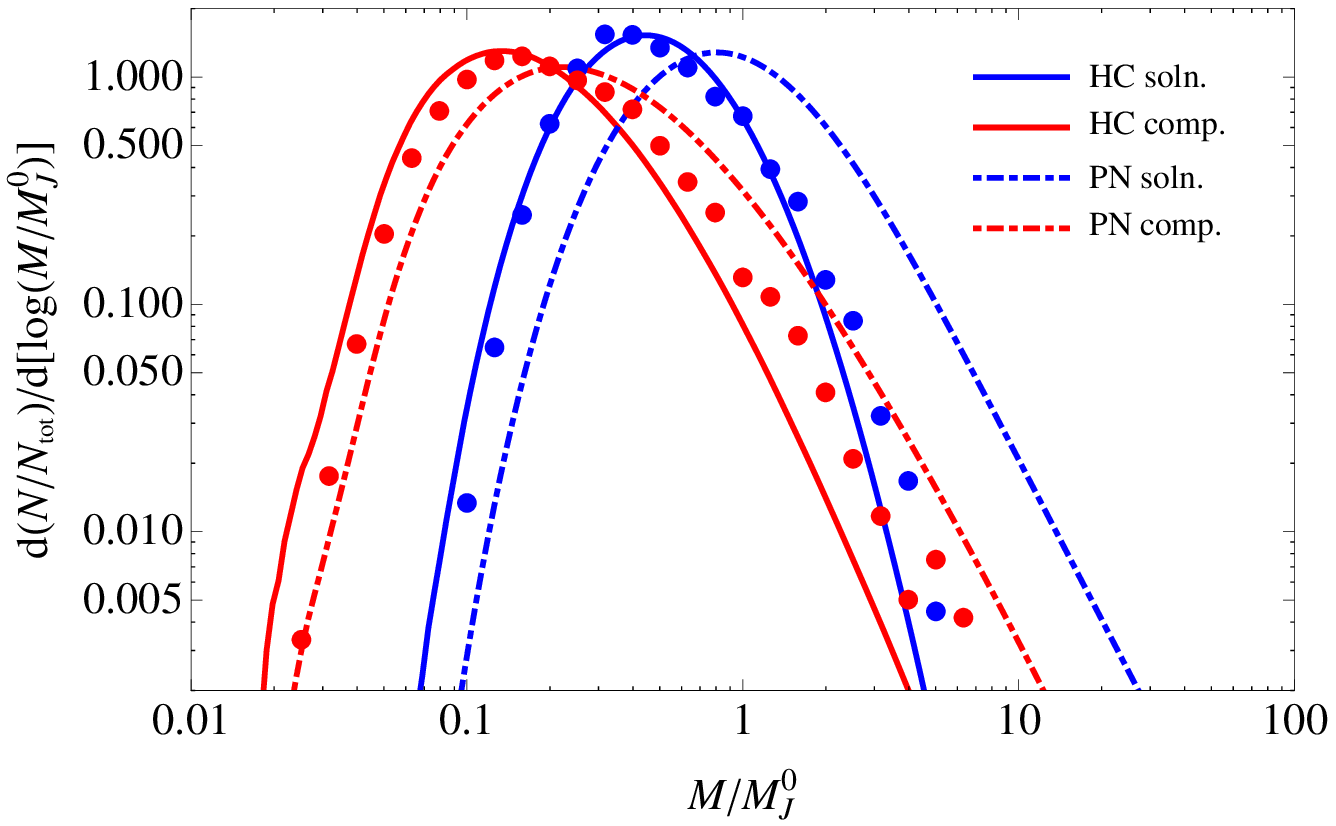}}\hfill
\subfigure[ turbulent support, $\lambda_{\mathrm{J}}^{0}/L=0.04$, $N_{\mathrm{BE}}=2\times 10^{5}$]{\includegraphics[width=0.49\linewidth]{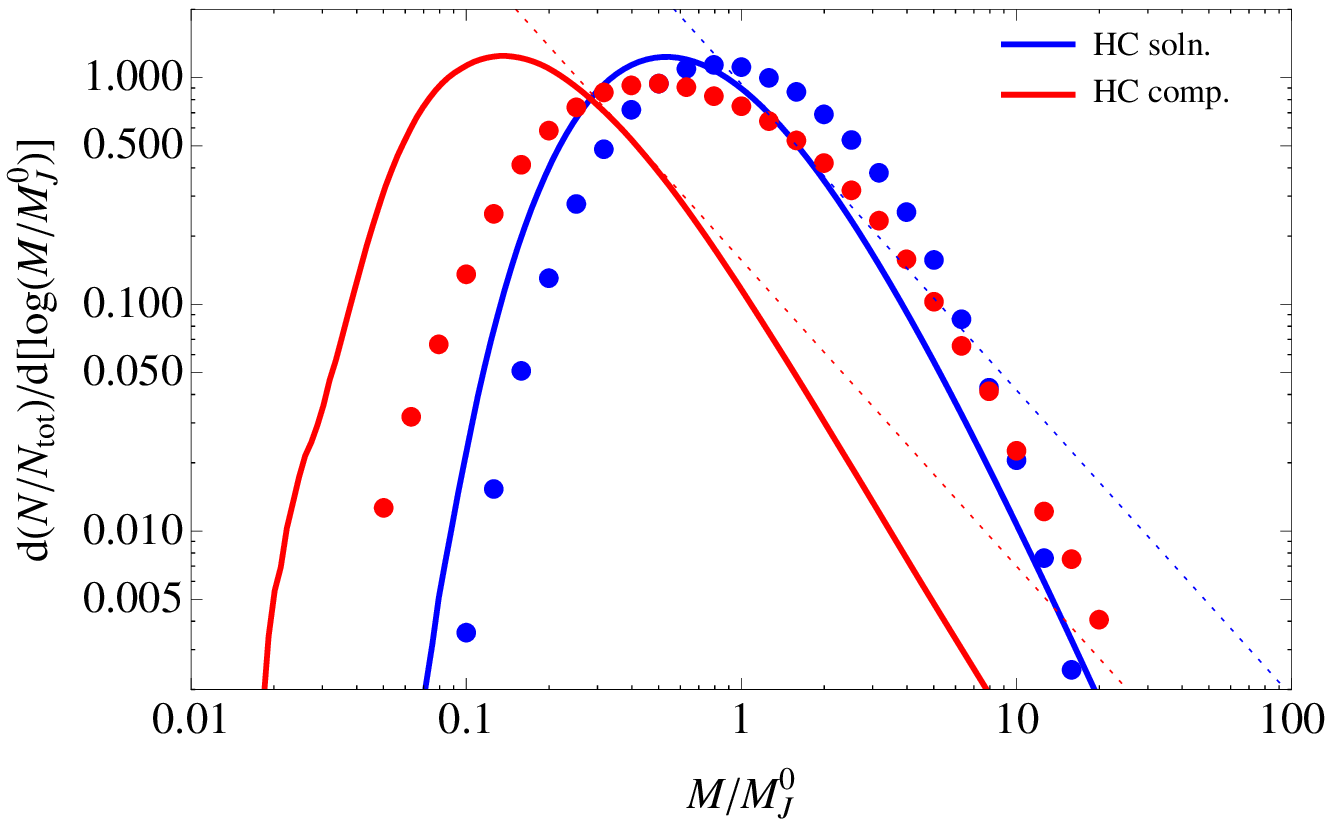}}
\subfigure[ turbulent support, $\lambda_{\mathrm{J}}^{0}/L=0.2$, $N_{\mathrm{BE}}=10^{3}$]{\includegraphics[width=0.49\linewidth]{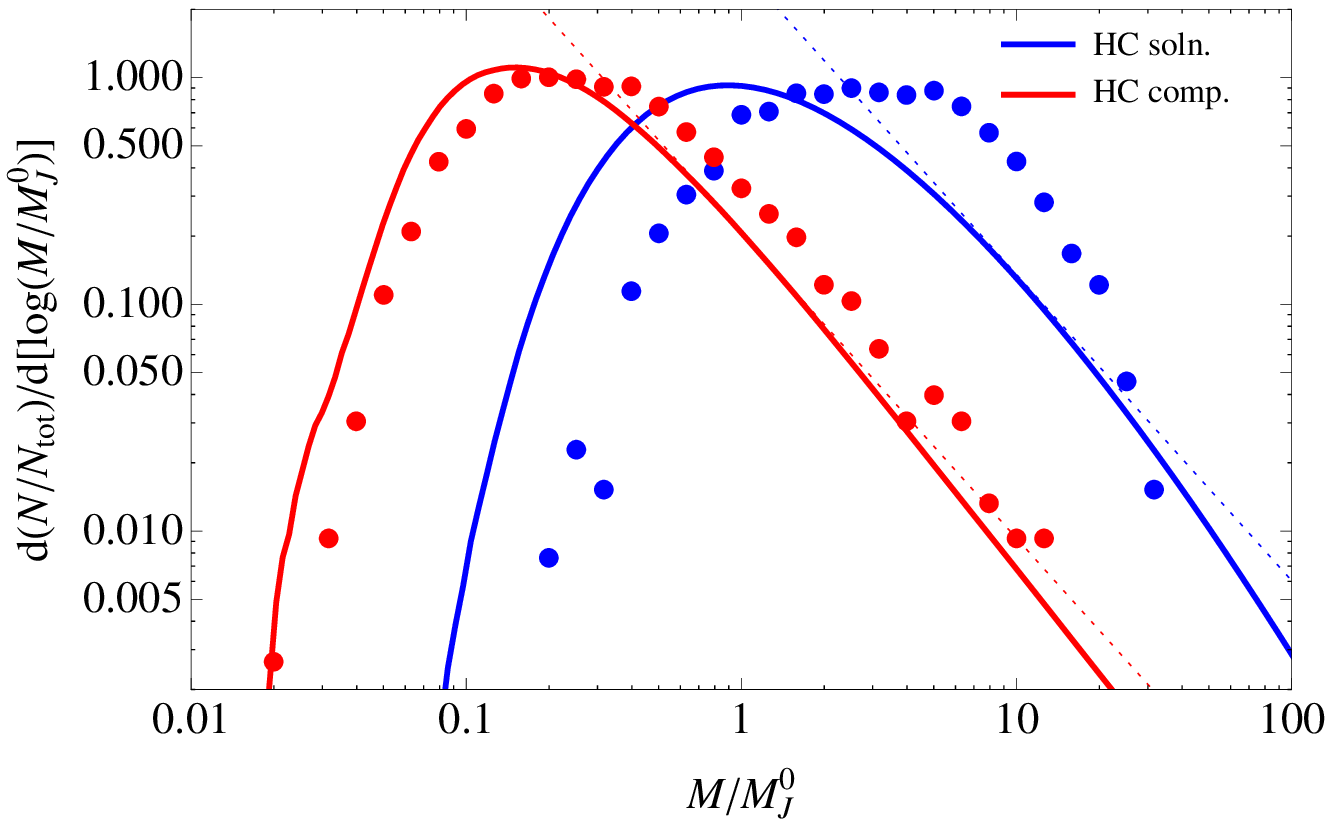}}\hfill
\caption{Comparison of the semi-analytic mass distribution following from the Hennebelle-Chabrier (HC) theory and the Padoan-Nordlund (PN) theory with the corresponding distributions obtained via clump-finding (large dots) for two different choices of global mass scale. The thin dotted lines are the tangents to the mass distributions with the Salpeter slope $x=1.35$. For an explanation of turbulent support, see Sect.~\ref{sc:hc_theory} and~\ref{sc:scaling}.
}
\label{fg:mass_spect+data}
\end{figure*}

\subsection{Purely thermal support}
\label{sc:therm}

Substitution of the time-averaged PDFs obtained from $1024^{3}$ simulations \citep{Federrath2008,FederDuv09} into equations~(\ref{eq:PN_mass_spect}) and~(\ref{eq:HC_mass_spect_log_therm}), yields the semi-analytic CMFs plotted in the top panels (a,b) of Fig.~\ref{fg:mass_spect+data}. These distributions are independent of the parameter $N_{\mathrm{BE}}$ because of the normalization by the total number of cores, $N_{\mathrm{tot}}$. As one can see, both the Padoan-Nordlund and the Hennebelle-Chabrier theory imply that the peak of the CMF is shifted towards lower masses for compressively driven turbulence. This is a direct consequence of the shape of the density PDFs.
Since compressive forcing produces larger density peaks, there is a higher fraction of small Jeans masses compared to solenoidal forcing. This results in the formation of smaller cores, as noted in Sect.~\ref{sc:num_res}).
As a consequence, the high-mass tail of the CMF~(\ref{eq:HC_mass_spect_log_therm}) is significantly less stiff in the compressive case (the asymptote for $\tilde{M}\gg 1$ is mainly given by the factor $\mathrm{pdf}(-2\log\tilde{M})$. 
The slope of the CMF~(\ref{eq:PN_mass_spect}), on the other hand, is determined by the factor $\tilde{M}^{-x}$, where $x$ is defined by equation~(\ref{eq:xslope}), in the limit of high masses.
The turbulence energy spectra computed from the simulation data yield the power-law exponents $\beta=1.86\pm0.05$ for solenoidal forcing and $\beta=1.94\pm0.05$ for compressive forcing \citep{FederDuv09}. Thus, the slopes of the high-mass tails predicted by the Padoan-Nordlund theory are $x\approx 2.3$ for solenoidal forcing and $x\approx 2.7$ for compressive forcing. 
 
The results from the clump-finding analysis are shown as dots in Fig.~\ref{fg:mass_spect+data} (a,b). Let us first consider the case $\lambda_{\mathrm{J}}^{0}/L=0.04$ ($N_{\mathrm{BE}}=2\times 10^{5}$). 
For compressively driven turbulence, the clump-finding data match the theoretical predictions very poorly.  As discussed in Section~\ref{sc:num_res}, the size of the smallest cores is less than the grid resolution in this case. Since the numerically unresolved cores are missing in the core statistics, 
the distribution is biased towards higher masses in comparison to the semi-analytic distributions.
In the case of solenoidal forcing, on the other hand, the smallest cores are at least marginally resolved (see Table~\ref{table:parameters}), and the low-mass wing as well as the peak position following from the clump-finding analysis are reasonably close to the theoretical predictions. While the distribution obtained by clump-finding roughly agrees with the CMF following from the Padoan-Nordlund theory also for $\tilde{M}\gg 1$, the slope of $x\approx 3.1$ (see Table~\ref{table:slopes}) is significantly steeper than the theoretical value $2.3$. In Sect~\ref{sc:comparison_turb}, it is shown that this discrepancy can be resolved by including turbulent pressure. Remarkably, the mass distributions from the clump-finding analysis are matched quite well by distributions of the form~(\ref{eq:HC_mass_spect_log_therm}) for $\lambda_{\mathrm{J}}^{0}/L=0.2$ ($N_{\mathrm{BE}}=10^{3}$). There are small deviations of the high-mass tails, which are, however, well within the error bars (see Fig~\ref{fg:dilsol1024}). In comparison to the distribution~({\ref{eq:PN_mass_spect}) predicted by the Padoan-Nordlund theory, large discrepancies become apparent. Since the cores are sufficiently resolved both for solenoidal and for compressive forcing for $\lambda_{\mathrm{J}}^{0}/L=0.2$ (see Table~\ref{table:parameters}), the most likely explanation is that $N_{\mathrm{BE}}$ is too small and, consequently, the high-mass cores are not within the asymptotic regime, for which the Padoan-Nordlund theory applies.

\begin{figure*}[t]
\centering
\subfigure[$\lambda_{\mathrm{J}}^{0}/L=0.04$, $N_{\mathrm{BE}}=2\times 10^{5}$]{\includegraphics[width=0.49\linewidth]{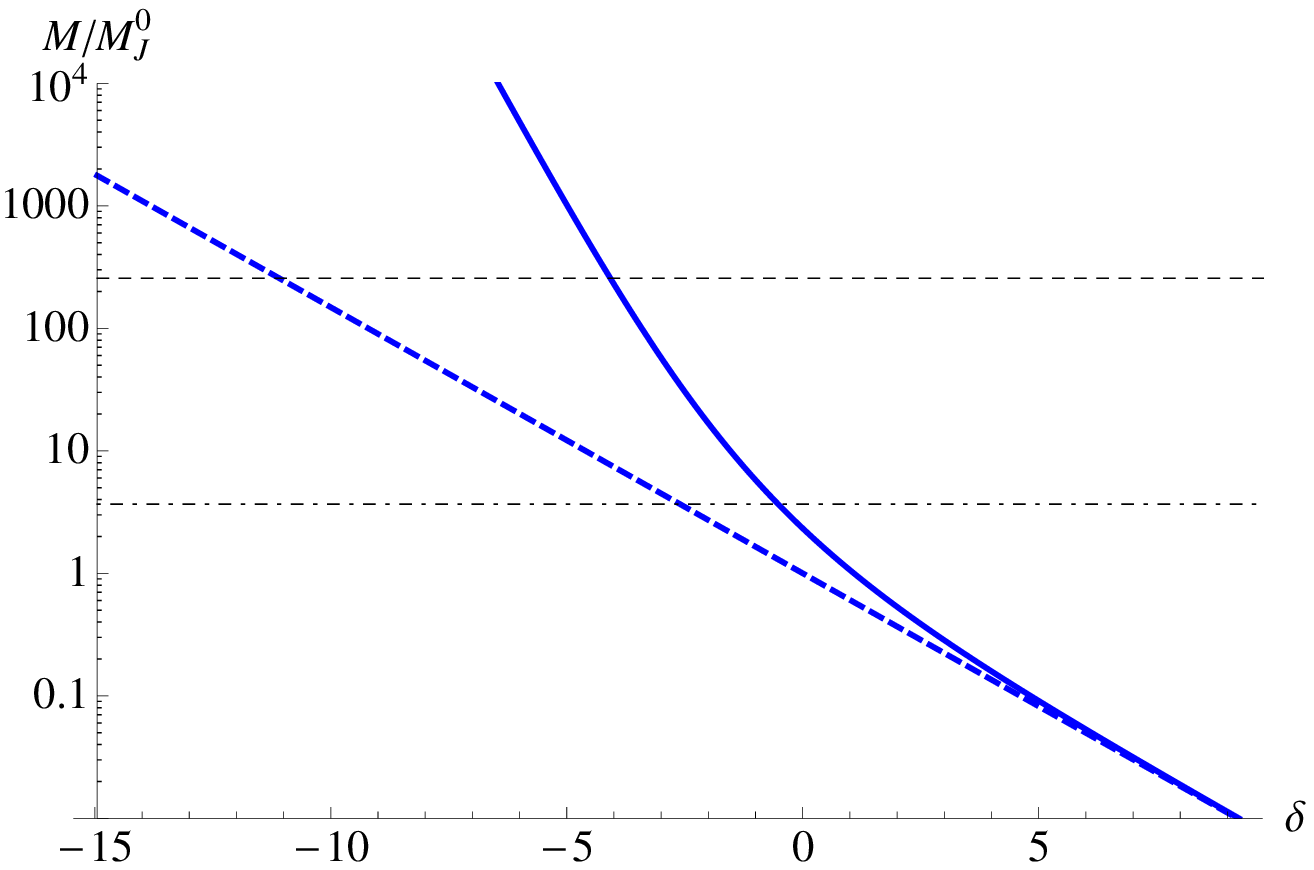}}
\subfigure[$\lambda_{\mathrm{J}}^{0}/L=0.2$, $N_{\mathrm{BE}}=10^{3}$]{\includegraphics[width=0.49\linewidth]{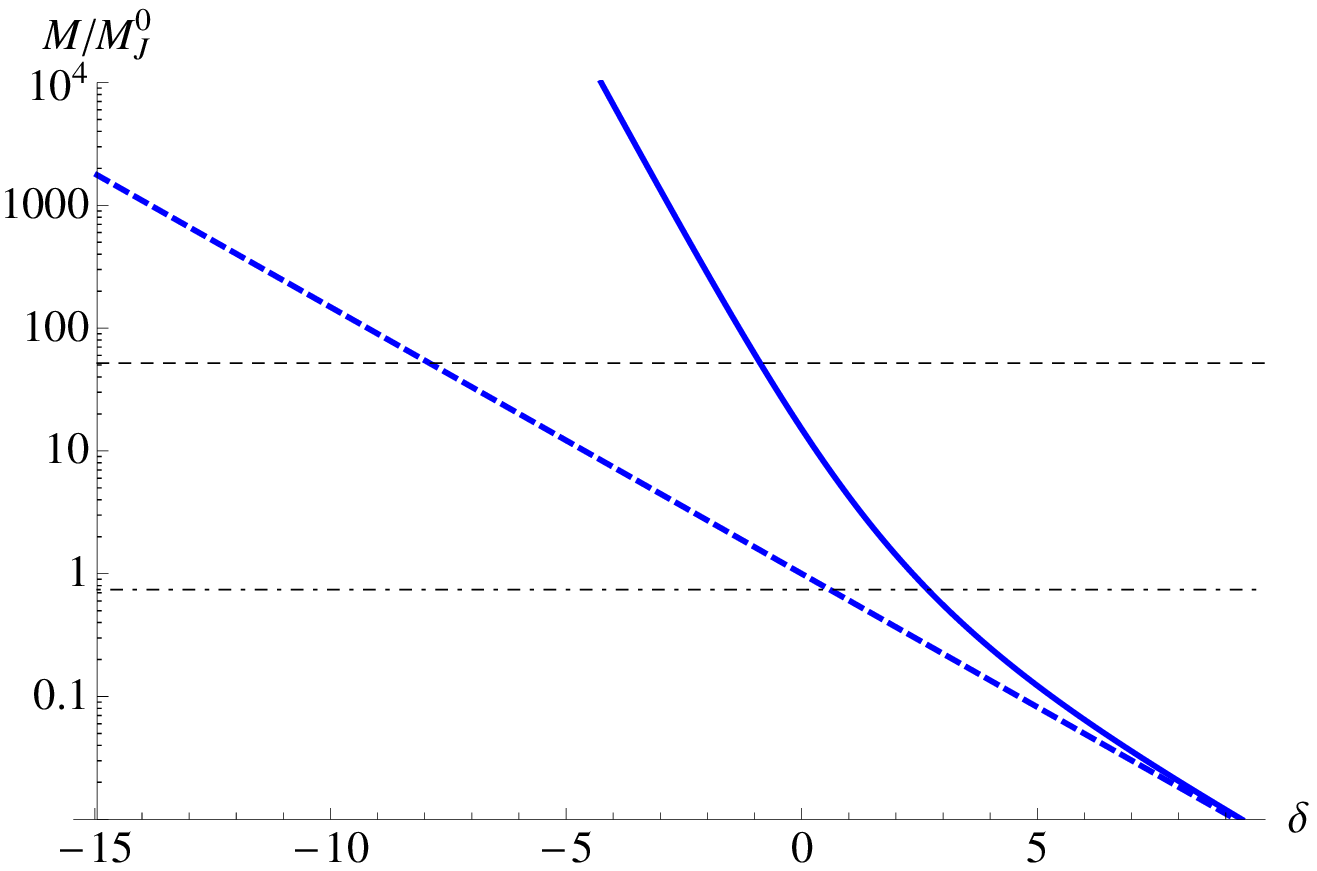}}\hfill
\caption{Normalised gravitationally unstable mass as a function of the logarithmic density fluctuation for purely thermal support (thick dashed lines) and with additional turbulent support (solid lines), as defined by equations~(\ref{eq:mass_dens_thermal}) and~(\ref{eq:mass_dens}). For solenoidal forcing (see Table~\ref{table:parameters}), the mass corresponding to the scale for which thermal pressure equals turbulent pressure ($\mathcal{M}_{\ast}\tilde{R}^{\eta}=1$) is indicated by the dot-dashed horizontal lines, and the dashed horizontal lines specify the mass for which the core scale equals the box size ($\tilde{R}=N_{\mathrm{BE}}^{1/3}$).}
\label{fg:mass-dens}
\end{figure*} 

\begin{table}[ht]
\caption{Dependence of various parameters on the mass scale and the forcing.}
\label{table:parameters}      
\centering                         
\begin{tabular}{lccllc}        
\hline                
\multicolumn{6}{c}{solenoidal forcing, $\mathcal{M}_{\mathrm{rms}}=5.3$, $\eta=0.43$}\\
\hline
$\lambda_{\mathrm{J}}^{0}/L$ & $N_{\mathrm{BE}}$ & $r_{\mathrm{min}}/\Delta$ &
 $\mathcal{M}_{\ast}$ & $\tilde{M}_{\mathrm{s}}$ & $\tilde{M}_{R=2L}$ \\  
\hline\hline                        
0.2 & $1.0\times 10^{3}$ & 4.6 & 1.53 & 0.74 & 52\\
0.04 & $1.2\times 10^{5}$ & 0.9 & 0.77 & 3.7 & 256\\
\hline                             
\multicolumn{6}{c}{compressive forcing, $\mathcal{M}_{\mathrm{rms}}=5.6$, $\eta=0.47$}\\
\hline
0.2 & $1.0\times 10^{3}$ & 1.0 & 1.52  & 0.82 & 57\\
0.04 & $1.2\times 10^{5}$ & 0.2 & 0.72 & 4.1 & 283\\
\hline                                   
\end{tabular}
\end{table}

\subsection{Turbulent support}
\label{sc:comparison_turb}

Since \citet{PadoanNordlund2002} considered only the thermal Jeans mass, we will concentrate on the Hennebelle-Chabrier theory for the case including both thermal and turbulent support.
Eliminating $\tilde{R}$ from equations~(\ref{eq:mass_dens}) by means of numerical root finding yields the mass-density relations plotted in Fig.~\ref{fg:mass-dens}. For comparison, also the relation~(\ref{eq:mass_dens_thermal}) for purely thermal support is shown. One can see that low-density cores of size $R$ greater than $\lambda_{\mathrm{J}}^{0}$ can maintain a much higher mass against gravitational collapse if turbulent pressure is included in the Jeans criterion. 
The high-density asymptote, on the other hand, coincides with the thermally supported branch, because cores of sufficiently high density are associated with small length scales $R$, for which the turbulent velocity dispersion becomes negligible compared to the speed of sound, i.~e., $\mathcal{M}_{\ast}\tilde{R}^{\eta}\ll 1$. This implies $\tilde{M}\simeq\tilde{R}\simeq \exp(-\delta/2)$. 
For the length scale $R_{\mathrm{s}}:=L(\mathcal{M}_{\mathrm{rms}}/\sqrt{3})^{-1/\eta}$, we have $\mathcal{M}_{\ast}\tilde{R}_{\mathrm{s}}^{\eta}=1$, i.~e., the turbulent pressure of the gas equals its thermal pressure. Substituting the values of the RMS Mach numbers and the scaling exponents from \citet{FederDuv09}, the values of $\mathcal{M}_{\mathrm{\ast}}$ listed in Table~\ref{table:parameters} are obtained. It follows that $R_{\mathrm{s}}\approx 0.074L$ and $0.082L$ for solenoidal and compressive forcing, respectively. These scales are close to the sonic length scales $\lambda_{\mathrm{s}}\approx 0.077L$ and $0.074L$, which \citet{FederDuv09} determined from the turbulence energy spectra. The corresponding dimensionless masses, $\tilde{M}_{\mathrm{s}}$, are also listed in Table~\ref{table:parameters}.\footnote{These values implicitly depend on the integral scale $L$, which is set to half the box size (see Sect.~\ref{sc:introduction}). More accurately, $L$ can be determined from the velocity structure functions (A. Kritsuk 2010, private communication). Indeed, this would entail a somewhat closer match between the semi-analytic core mass distributions and the clump-finding data. Since the trends remain unaltered, however, we keep the simple definition of $L$ in this article.}

An important limitation of calculations based on the approximation~(\ref{eq:HC_mass_spect}) is that $R$ has to be small compared to the box size $2L$, i.~e., $\tilde{R}\ll N_{\mathrm{BE}}^{1/3}$. The mass parameters $\tilde{M}_{R=2L}$ corresponding to $\tilde{R}=N_{\mathrm{BE}}^{1/3}$ are indicated by the dashed horizontal lines in Fig.~\ref{fg:mass-dens}. The values of $\tilde{M}_{R=2L}$ are listed in Table~\ref{table:parameters} for all parameter sets. As one can see, we have the constraints $\tilde{M}\ll 50$ and $\tilde{M}\ll 250$ for $\lambda_{\mathrm{J}}^{0}/L=0.2$ and $0.04$, respectively. Since these limits are much higher than the peak positions of the mass distributions, there is a sufficient margin to study the high-mass wings. 

The PDF data for the density fluctuation $\delta$ yield the mass distributions plotted in the bottom panels (c,d) of Fig.~\ref{fg:mass_spect+data}.
Comparing the mass distributions with and without turbulent support, the high-mass tails are flatter in the former case. As expected, the difference is more pronounced for $\lambda_{\mathrm{J}}^{0}/L=0.2$, because of the higher contribution of turbulent pressure for large core masses
(see Fig.~\ref{fg:mass-dens}). Remarkably, the tail of the mass distribution for compressively driven turbulence that is plotted in Fig.~\ref{fg:mass_spect+data} (d) is much less stiff than what PN07 reported for hydrodynamic turbulence and is in very good agreement
with the Salpeter power law.
The peak positions, on the other hand, are nearly unaffected, because the turbulent pressure on the
length scales corresponding to the peaks of the mass distributions is small compared to the thermal pressure.

Regarding the mass distributions obtained by clump-finding with turbulent support (large dots in the bottom panels of Fig~\ref{fg:mass_spect+data}), similar trends as for the distributions with purely thermal support can be seen. 
For $\lambda_{\mathrm{J}}^{0}/L=0.04$ ($N_{\mathrm{BE}}\approx 1.2\times 10^{5}$), the clump-finding distribution is shifted towards higher masses for compressively driven turbulence because the smallest cores cannot be resolved (see Sect.~\ref{sc:therm}). The discrepancy
is much smaller in the case of solenoidal forcing, for which the overall shape of the semi-analytic
and clump-finding distributions agree quite well. However, the slopes of the tails following from the
clump-finding data are steeper in both cases. Consequently, it appears that the clump-finding algorithm
underestimates the turbulent support of high-mass cores, provided that the theoretical description
of the CMF is correct. 

In the case $\lambda_{\mathrm{J}}^{0}/L=0.2$ ($N_{\mathrm{BE}}=10^{3}$), the mass distributions agree well for compressive forcing (except for a small shift), whereas the mass distribution obtained by clump-finding is markedly different from the semi-analytic model for solenoidal forcing. 
The analysis in Sect~\ref{sc:num_res} showed that the cores tend to be smaller for compressively driven turbulence. In accordance with the clump-finding results, the Hennebelle-Chabrier theory implies a significant flattening of the high-mass wing in this case, and the slope is found to be close to the Salpeter value. For solenoidal forcing, however, a significant number of cores have sizes that are of the same order of magnitude as the integral scale $L$ (see Table~\ref{table:sizes_summary} and Fig.~\ref{fg:resolution_sizes2}). 
This entails a violation of the basic assumption that was made in the derivation of equation~(\ref{eq:HC_mass_spect_log}). Apart from that, it is important to realize that both the clump-finding algorithm and the semi-analytic approach rely on the notions of the Bonnor-Ebert or Jeans mass and turbulence pressure. Being based on the collapse of a spherical cloud, the former is highly idealised and does not apply to the fully non-linear regime, while turbulence pressure is an ensemble-average property of isotropic turbulence.  A mechanism that might account for the massive cores is merging. Indeed, since the gas is less compressed, the cores are more extended and should merge more easily than in the compressible forcing case. Also they should live longer, as it takes more time for them to be assembled, and it is likely for these cores to be destroyed by violent passing shocks. 

\section{Conclusions} \label{sc:conclusion}

Computing the distributions of gravitationally unstable cores by means of a clump-finding algorithm as well as semi-analytic methods for hydrodynamic simulations of forced supersonic turbulence, we have found significant differences resulting from the turbulence forcing, the choice of the global mass scale, and the effects of turbulent support. For a comparison with theoretical predictions, the numerical probability density functions of the mass density fluctuations were used to evaluate the formulae for the mass distribution.
Our analysis completes a comprehensive study of the influence of forcing on the density and velocity statistics of isothermal supersonic turbulence \citep{Federrath2008,SchmFeder08b,FederDuv09,Federrath2009}. In the following, we summarize the main results:

\begin{enumerate}

\item For hydrodynamic simulations without explicit treatment of self-gravity, the CMF depends on the choice of the global mass scale, which determines the number of Bonnor-Ebert masses, $N_{\mathrm{BE}}$, with respect to the mean density in the computational domain. For clump-finding, one has to observe two constraints: $N_{\mathrm{BE}}$ must be sufficiently large to allow for many cores, but if $N_{\mathrm{BE}}$ is chosen too large, a significant fraction of cores will be numerically unresolved and the resulting CMF will be shifted towards higher masses. In our analysis, these constraints are well satisfied for two cases: Firstly, solenoidal forcing with $N_{\mathrm{BE}}\approx 1.2\times 10^{5}$ and, secondly, compressive forcing with $N_{\mathrm{BE}}=10^{3}$. The lower value of $N_{\mathrm{BE}}$ is consistent with Larson-type relations.

\item Comparing the results for solenoidal and compressive forcing, the most noticeable trends are that the CMF for compressively driven turbulence displays more low-mass cores, while the high-mass tails are flatter, particularly, if turbulent support is significant. These trends are implied by equations~(\ref{eq:mass_dens_thermal}) and~(\ref{eq:mass_dens}) for the larger density contrast produced by compressive forcing (also see Fig.~\ref{fg:mass-dens}). We also found that the CMF peaks at lower mass in the case of compressive forcing. These results follow both from the clump-finding analysis and the semi-analytic computation of the mass distributions.

\item Because of the considerable scatter of core masses and possible biases of the clump-finding algorithm, it is difficult to ascertain power laws for the mass distributions. Nevertheless, we attempted to determine power-law exponents for the high-mass tails from least square fits and by means of MLE. The results agree within the statistical uncertainties for both methods. However, the results for purely thermal support are at odds with the theory of \citet{PadoanNordlund2002}, which asymptotically applies to $N_{\mathrm{BE}}\rightarrow\infty$ and predicts power-law tails for the CMF. In particular, the difference between the exponents following from the clump-finding analysis is more pronounced than the theoretical prediction.

\item The mass distributions for cores with purely thermal support agree well with the theory of \cite{HennebelleChabrier2008} for $N_{\mathrm{BE}}=10^{3}$, although the tails are less stiff towards high masses than what is expected on the basis of the semi-analytic models.
There are large discrepancies for $N_{\mathrm{BE}}=1.2\times 10^{5}$, but the disagreement might be spurious because of the strong dependence of the high-mass wings on the numerical resolution. Apparently, the mass distributions for high values of $N_{\mathrm{BE}}$ are in closer agreement with \citet{PadoanNordlund2002} if purely thermal support is assumed. However, it should be noted that the mass-density pdfs will be stronlgy affected by self-gravity in this case. 

\item For $N_{\mathrm{BE}}=10^{3}$, turbulent pressure is important for a wider range of core masses than in the case $N_{\mathrm{BE}}=1.2\times 10^{5}$.
For this reason, the CMFs change substantially if turbulent support is included. The effect is particularly strong for
turbulence that is driven by solenoidal forcing, because the mass and the size of virtually all cores is increased by turbulent support. However, the mass distribution obtained by clump-finding cannot be reproduced theoretically, because the basic assumption
that the size of the cores is much smaller than the integral scale is clearly not fulfilled in this case. Moreover, a stability criterion that is based on the notions of Jeans mass and turbulent pressure might fail if the cores become too large. On the other hand, we find very good agreement between the clump-finding analysis and the Hennebelle-Chabrier for compressively driven turbulence with $N_{\mathrm{BE}}=10^{3}$. Compared to purely thermal support, the high-mass tail is considerably flatter in this
case.

\end{enumerate}

One of the difficulties in assessing the applicability of the semi-analytic models is that we cannot disentangle limitations of the underlying theoretical concepts from shortcomings of the clump-finding algorithm. Studying the influence of the clump-finding parameters, it becomes clear that the shape of the resulting mass distribution varies significantly with these parameters. While the peak position appears to be quite robust, the tails are more affected by the tuning of the clump-finding algorithm. Apart from these uncertainties and the large scatter of the core masses, there might be systematic biases. In particular, it is not entirely clear whether clump-finding traces down the smallest gravitationally unstable cores.

Even so, we have been able to shed more light on the gravitational fragmentation of turbulent gas. The influence of the forcing is irrefutable for a range of length scales that certainly extends beyond the energy-containing range. An open question is which regime in the interstellar medium is suitably modelled by a particular mode of forcing. If compressive excitation of turbulence occurs on length scales that tend to be larger than the size of molecular clouds, then the core statistics on much smaller scales, i.~e., inside the cores of the clouds, still might be universal. \citet{FederDuv09}, however, suggested that different forcing mechanisms can produce genuine differences in molecular clouds, which affect the properties of turbulence even on the length scales of molecular cloud cores. This calls for further advances both on the observational and the theoretical side. 

A further important result is the influence of turbulent pressure on the gas fragmentation. From the theoretical point of view this is not new at all. But we have been able to demonstrate numerically that turbulent support entails a flattening of the CMF towards high masses. In one extreme case, the slope of the high-mass tail was found to be close to the Salpeter slope. In this respect, the effect of turbulent pressure on gravitational fragmentation is analogous to the effect of magnetic pressure in magnetohydrodynamical turbulence. This can be understood on the basis of the dispersion relation resulting from a linear stability analysis, which shows that the turbulent pressure as well as the magnetic pressure contribute to the effective pressure of the gas. We do not suggest that the observed CMF might be explainable in terms of hydrodynamical turbulence only, because it is known that magnetic fields play an important role in the interstellar medium. However, our results show that turbulent pressure might significantly contribute to the slope of the high-mass tail of the CMF. Apart from that, this effect is important in numerical simulations, where unresolved turbulent velocity fluctuations are of the order of the speed of sound or higher. In this case, the corresponding turbulent pressure can be computed with a subgrid-scale model \citep{Schm09}.

In order to achieve further progress with the theoretical explanation of the CMF, it is paramount to account for processes that modify the PDFs of the mass density. One such process is the thermal instability induced by radiative cooling in the interstellar medium. It was already noticed by \citet{PassVaz98} that polytropic equations of state, which are simple models for cooling, lead to power-law tails in the density PDFs. This was confirmed by
\citet{LiKless03} for three-dimensional turbulence-in-a-box simulations and by \citet{AudHenne09} for three-dimensional simulations of colliding flows with a polytropic exponent $\gamma=0.7$, although a power law was not obtained for thermally bistable turbulence. \citet{SeiSchm09}, on the other hand, find power-law tails for thermally bistable turbulence produced by compressive forcing, using the cooling function of \citet{AudHenne05}. The effects of a polytropic equation of state on the CMF were analysed by \citet{HenneChab09}. 

Another process that gives rise to density PDFs with power-law tails is the accretion and contraction of gas in self-gravitating turbulence \citep[e.~g.][]{KlessHei00,FederGlov08}. Moreover, dense structures can merge due to their mutual gravitational attraction. Consequently, the statistical characteristics will evolve with time \citep{Kless00}, and the corresponding core mass distribution can differ significantly from the non-self-gravitating case \citep{Kless01}, depending on the relative strength of self-gravity and the evolutionary stage of the star-forming cloud. Indeed, power-law high-density tails are observed in high-resolution extinction maps of nearby molecular clouds \citep{KainBeu09}.

In numerical simulations of self-gravitating turbulence, the mass distribution of gravitationally collapsing objects can be determined directly. However, the sink particles that are usually applied to numerically capture the collapsing gas \citep{BateBon95,KrumMcKee04,JappKless05,PadNord09,FederBan09} do not directly correspond to the cores we have considered here. While sink particles are dynamic objects that accrete gas and thus can acquire different masses, they are obviously not associated with a variable length scale, as is the case for cores\footnote{There is a fixed accretion radius extending over a few grid cells, which can be interpreted as the length scale associated with sink particles.}. Theoretically, the relation between length scale and mass becomes apparent in the Hennebelle-Chabrier theory, but also the clump-finding algorithm identifies regions of variable size as cores. For this reason, relating the mass-distributions of sink particles and cores is non-trivial. Nevertheless, the approach via sink particles has the considerable advantage that concepts such as the Bonner-Ebert mass, which is rather arbitrary in the highly non-linear regime, are not required. In any case, including self-gravity in turbulence simulations is indispensable to improve our understanding of gravoturbulent fragmentation.

\begin{acknowledgements}
We thank Paolo Padoan (UCSD) for making his clump-finding algorithm available for this work and for critical comments. We are
grateful to Patrick Hennebelle (ENS Paris) for numerous discussions and plenty of advice. Thanks also to Jens C. Niemeyer (IAG) for his support. The simulations and data analysis used resources from HLRBII project h0972 at the Leibniz Supercomputer Centre in Garching, Germany. CF is grateful for financial support from the International Max Planck Research School for Astronomy and Cosmic Physics (\textsc{imprs-a}) and the Heidelberg Graduate School of Fundamental Physics (\textsc{hgsfp}), which is funded by the Excellence Initiative of the Deutsche Forschungsgemeinschaft (\textsc{dfg}) \textsc{gsc} 129/1. RSK thanks for financial resources provided by the {\em Deutsche Forschungsgemeinschaft} under grants no.\ KL 1358/1, KL 1358/4, KL 1359/5. RSK furthermore thanks for subsidies from a Frontier grant of Heidelberg University sponsored by the German Excellence Initiative and for support from the {\em Landesstiftung Baden-W\"urttemberg} via their program International Collaboration II, grant no.\ P-LS-SPII/18. RSK also acknowledges financial support from the German {\em Bundesministerium f\"ur Bildung und Forschung} via the ASTRONET project STAR FORMAT (grant 05A09VHA).
\end{acknowledgements}

\bibliographystyle{aa}
\bibliography{CoreTurb}

\appendix
\section{Maximum likelihood estimation for power law scaling parameters}\label{ap:mlemethod}
Estimating the parameters $\alpha$ and $x_{\mathrm{min}}$ of a power law distribution of the form 
\begin{equation}\label{eq:powerlaw}
d(x)=C x^{-\alpha},\;\;\;x\geq x_{\mathrm{min}} 
\end{equation}
with a constant $C$ and exponent $\alpha$ by simply fitting straight lines on a log-log plot introduces large biases \citep[e.g.][]{Newman2005,Clauset2007}. Furthermore, this technique gives no information whether assuming a power law is reasonable or not. In order to accurately estimate the scaling parameters of the high-mass range in our mass distributions we followed the MLE method of \citet{Clauset2007}. In this appendix we give a brief overview of the method. For a detailed description please refer to the original paper.
 
\citet{Clauset2007} use a maximum likelihood estimator 
\begin{equation}\label{eq:mle}
\hat{\alpha}=1+n\left[\sum_{i=1}^{n}{\ln \frac{x_{i}}{x_{\mathrm{min}}}}\right]^{-1}
\end{equation}
for the power law exponent $\alpha$ assuming that the dataset, consisting of $n$ values $x_{i}$ with $i=1 \ldots n$, is drawn from a continuous power law distribution for $x_{\mathrm{i}}\geq x_{\mathrm{min}}$. $\hat{\alpha}$ is the estimated value for  $\alpha$ of the underlying distribution (eqn.~\ref{eq:powerlaw}). The value of the lower bound $x_{\mathrm{min}}$ of the distribution is the crucial factor in this calculation. If the estimated $\hat{x}_{\mathrm{min}}$ is too small ($\hat{x}_{\mathrm{min}} < x_{\mathrm{min}}$) one will fit a power-law to non-power-law data and the parameter estimation will be biased. A too high $\hat{x}_{\mathrm{min}} > x_{\mathrm{min}}$ value will throw away a certain amount of legitimate, power-law-distributed data points. A reasonable value for $\hat{x}_{\mathrm{min}}$ minimises the Kolmogorov-Smirnov (KS) statistics defined by

\begin{equation}\label{eq:KSstatistic}
D=\underset{x \geq x_{\mathrm{min}}}{\mathrm{max}}\ |S(x)-P(x)|,
\end{equation}

\noindent where $S(x)$ is the cumulative distribution function (CDF) of the observed data points and $P(x)$ is the CDF of the fitted power law model. The optimal value for $\hat{x}_{\mathrm{min}}$ gives the minimum difference between the data and the fitted model which is just the minimised KS-statistics $D$ of eqn.~(\ref{eq:KSstatistic}).

To check if the observed dataset is likely to be drawn from a power law distribution we calculate a $p$-value using a semi-parametric bootstrap approach. Therefore, we create $2500$ synthetic datasets per snapshot which follow a real power law with the estimated parameters $\hat{x}_{\mathrm{min}}$ and $\hat{\alpha}$ of the observed core mass values. For $x_{\mathrm{i}} < \hat{x}_{\mathrm{min}}$ the synthetic distributions follow the same non-power law behaviour as the observed dataset.
For each of the synthetic distributions we again estimate the parameters $\alpha$ and $x_{\mathrm{min}}$ and compute the KS-statistics.
The $p$-value is the fraction of the KS-statistics of the synthetic datasets whose value is higher than the KS-statistics of the observed dataset.That means that $p$ is the probability of getting a goodness of fit, e.g. the KS-statistics, for a real power law distributed synthetic dataset which is at least as bad as the goodness of fit of our observed dataset.
For $p \leq 0.1$ the power law assumption has to be ruled out. Note that a $p$-value higher than $0.1$ does not necessarily mean that the underlying distribution follows a power law. The $p$-value can only rule out the hypothesis of a power law distribution. 

\end{document}